\documentclass{emulateapj}
\usepackage{longtable}

\newcommand{\HI}{H$\;${\small\rm I}\relax}
\newcommand{\hi}{H$\;${\small\rm I}\relax}
\newcommand{\hii}{H$\;${\small\rm II}\relax}

\newcommand{\mgii}{Mg$\;${\small\rm II}\relax}
\newcommand{\feii}{Fe$\;${\small\rm II}\relax}
\newcommand{\civ}{C$\;${\small\rm IV}\relax}

\newcommand{\kms}{km~s$^{-1}$\relax}

\newcommand{\gammaone}{1.19 $\pm$ 0.56}
\newcommand{\gammatwo}{1.33 $\pm$ 0.61}
\newcommand{\None}{0.85}
\newcommand{\Ntwo}{0.59}
\newcommand{\gammac}{1.83 $\pm$ 0.21}
\newcommand{\Nc}{1.62} 

\slugcomment{Accepted for publication in ApJ}
\shortauthors{Ribaudo, Lehner, \& Howk}
\shorttitle{HST Study of LLSs}

\begin{document}

\title{A {\it Hubble Space Telescope} Study of Lyman Limit Systems: Census and Evolution\altaffilmark{1}}

\author{ Joseph Ribaudo, Nicolas Lehner, J. Christopher Howk
}
\affil{Department of Physics, University of Notre Dame,  Notre Dame, IN 46556}

\altaffiltext{1}{Based on observations made with the NASA/ESA Hubble Space Telescope, obtained at the Space Telescope Science Institute, which is operated by the Association of Universities for Research in Astronomy, Inc. under NASA contract No. NAS5-26555.}

%%%%%%%%%%%%%%%%%%%%%%%%%%%%%%%%%%%%%%%%%%%%%%%%%%%%%%%%%%%%%%%%%%%%%%%

\begin{abstract}

We present a survey for optically thick Lyman limit absorbers at
$z<2.6$  using
archival {\em Hubble Space Telescope} observations with the Faint
Object Spectrograph and Space Telescope Imaging Spectrograph.  We
identify 206 Lyman limit systems (LLSs) increasing the number
of catalogued LLSs at $z<2.6$ by a factor of $\sim$10. We compile a
statistical sample of $50$ $\tau_{\rm LLS} \geq 2$ LLSs drawn from 249
QSO sight lines that avoid known targeting biases.  The incidence of
such LLSs per unit redshift, $l(z)=dn/dz$, at these redshifts
 is well described by a single power law, $l(z) \propto (1+z)^\gamma$, 
with $\gamma=$\gammatwo\ at $z<2.6$, or with $\gamma=$\,\gammac\  over
the redshift range $0.2 \leq z \leq 4.9$. The incidence of LLSs per
absorption distance, $l(X)$, decreases by a factor of $\sim$1.5 over
the $\sim$0.6 Gyr from $z=4.9$ to 3.5; $l(X)$ evolves much more
slowly at low redshifts, decreasing by a similar factor over the
$\sim$8 Gyr from $z=2.6$ to 0.25.  We show that the column density
distribution function, $f(N_{\rm HI})$, at low redshift is not well
fitted by a single power law index ($f(N_{\rm HI}) \propto N^{-
\beta}_{\rm HI}$) over the column density range $13 \le \log N_{\rm
HI} \le 22$ or $\log N_{\rm HI} \ge 17.2$.  While
low and high redshift $f(N_{\rm HI})$ distributions are
consistent for $\log N_{\rm HI}>19.0$, there is some evidence
that $f(N_{\rm HI})$ evolves with $z$ for $\log N_{\rm HI}\la 17.7$,
possibly due to the evolution of the UV background and galactic
feedback.   Assuming LLSs are associated with
individual galaxies, we show that the physical cross section of the
optically thick envelopes of galaxies decreased by a factor of $\sim$9
from $z\sim$5 to 2 and has remained relatively constant since
that time.  We argue that a significant fraction of the observed
population of LLSs arises in the circumgalactic gas of sub-$L_*$ galaxies.

\end{abstract}
\keywords{Intergactic Medium --- Galaxies: Quasars: Absorption lines}

%%%%%%%%%%%%%%%%%%%%%%%%%%%%%%%%%%%%%%%%%%%%%%%%%%%%%%%%%%%%%%%%%%%%%%%

\section{Introduction}
\label{sec:intro}

The absorption features seen in the spectra of QSOs provide a unique 
opportunity to probe the intergalactic and galactic regions which intersect the lines of sight.
In particular, \HI\ absorption studies have allowed us to examine the distribution of gas associated 
with galaxies, the intergalactic medium (IGM), and the extended 
gaseous regions of galaxies which serve as an interface to the IGM, over the majority of cosmic time.  
Often these \HI\ absorbers are placed in three general categories
dependent on the \HI\ column density ($N_{\rm H I}$) of the absorber. The low column density Lyman-$\alpha$ 
forest absorbers ($N_{\rm H I} < 10^{16}$ cm$^{-2}$)  
are associated with the diffuse IGM \citep[see review by][]{rauch98}. These systems probe low-density, 
highly ionized gas and are thought to trace the dark matter distribution throughout the IGM \citep{jena05} 
as well as contain the bulk of the baryons at high redshift \citep{miralda96} and a significant amount of the baryons even 
today \citep[e.g.,][]{penton04,lehner07,danforth08}. At the other end, the high column
density damped Lyman-$\alpha$ absorbers (DLAs, $N_{\rm H I} > 10^{20.3}$ cm$^{-2}$) appear associated with the main 
bodies of galaxies (see review by Wolfe (2005), although see Rauch et al. (2009)). These high-density, 
predominantly neutral systems serve as neutral gas reservoirs for high redshift star formation \citep{pw09}. 

The intermediate column density systems mark the transition from the optically thin Lyman-$\alpha$ forest to the 
optically thick absorbers found in and around the extended
regions of galaxies. Typically these absorbers are easy to
identify in QSO spectra due to the characteristic attenuation of QSO flux by the Lyman limit at $\sim$912 \AA\ in 
the rest frame. These intermediate column
density systems are segmented into three additional categories. The low column density absorbers 
($10^{16}$ cm$^{-2} \leq N_{\rm H I} < 10^{17.2}$ cm$^{-2}$) are known as partial Lyman limit systems (PLLSs), the 
intermediate column density absorbers ($10^{17.2}$ cm$^{-2} \leq N_{\rm H I} < 10^{19}$ cm$^{-2}$)
are known simply as Lyman limit systems \citep[LLSs,][]{tytler82}, and the high column density absorbers 
($10^{19}$ cm$^{-2} \leq N_{\rm H I} <10^{20.3}$ cm$^{-2}$) are known as super 
Lyman limit systems \citep[SLLSs, a.k.a. sub-DLAs;][]{omeara07,peroux02,kulkarni07}. These absorbers are the 
least well-studied and physically understood class of absorbers, especially at $z \la 2.6$, 
i.e. over the past $\sim$10 Gyr of cosmic time. The reason for that is because at redshifts $z \la 2.6$, the Lyman 
limit is shifted into the UV, requiring the need for space-based UV observations to observe the 
Lyman break in spectra. 

To date, the majority of spectra used in LLS surveys have been taken from ground-based observations, 
providing an adequate statistical description of the high redshift ($z \ga 3.0$) absorbers, most recently by \citet{prochaska10} 
and \citet{sc10}.  Previous and recent surveys that
partially probe the $z<2.6$ regime \citep{tytler82,sargent89,lanzetta91,storrie94,stengler95, sc10} have produced samples of tens of 
LLSs spanning the redshift range $0 \la z \la 4$. 
These surveys studied the statistical nature of LLSs, with some conflicting conclusions as to the evolution 
of these absorbers over cosmic time. A complete understanding
of these optically thick absorbers is crucial as these systems in part determine the strength and shape of the ionizing 
ultraviolet background \citep[UVB,][]{shull99,hm96,zp93}. 
Due to the position of LLS column density with respect to Lyman-$\alpha$ forest systems and DLAs, a priori it is natural to view LLSs
as tracing the IGM/galaxy interface. Thus they may provide a potentially unique probe of material moving in and out of galaxies over time.
 It is for these reasons that the incidence of optically thick
absorbers as a function of redshift and the frequency distribution of absorbers as a function of $N_{\rm HI}$ 
serve as a critical parameter in modern cosmological simulations 
\citep{rauch98,keres05,keres09,kh09,nagamine10}.

Observations have linked LLSs to
the extended regions of galaxies, including their gaseous halos, winds, and the interactions of these with the 
IGM \citep[e.g.,][]{simcoe06, prochaska04,prochaska06,lehner09,stocke10}. 
Simulations have 
also shown a physical connection between LLSs and galaxies of a wide range of masses at $z\sim$2 to 4 \citep{gardner01,kg07}. 
In addition, surveys of \mgii\ and \civ\ absorbers have shown connections 
to extended galactic environments and indicate the metal absorbers trace similar physical regions as LLSs 
\citep[e.g.,][]{chen01,chen10,church00,church05,cc98,steidel92}. 
\mgii\ absorbers have been studied extensively in optical surveys where the absorbers are observed over the 
redshift range $0.3 \leq z \leq 2.2$  and led the way in connecting
QSO absorption features with galactic environments \citep[e.g.,][]{tytler87,
petitjean90,nestor05}. Due to the nature of the \mgii\ absorption lines, which are strong and easily saturated, 
measurements of the \mgii\ column density are often impossible. This limits
the information available as to their 
origin, metallicity, and physical properties. LLSs provide a complementary approach in understanding the gas 
around galaxies and provide a reliable estimate of $N_{\rm HI}$
for $\tau_{\rm LLS} \le 2.5$ (from the Lyman limit) and $\tau_{\rm LLS} \ge 50$ (from the Lyman-$\alpha$ line) 
absorbers. For example, measurements of $N_{\rm HI}$ allow an examination of the frequency distribution with 
column density, which provides additional insight into the evolution of the 
strength and shape of the UVB over cosmic time.
   
In this work we analyze the population of LLSs at low redshift using a new sample of spectra from archival  \textit{Hubble Space Telescope (HST)} 
observations with the 
Faint Object Spectrograph (FOS) and the Space Telescope Imaging Spectrograph (STIS). We present the most complete survey to date of 
LLSs at $z \le 2.6$. We catalogue 206 LLSs at $z < 2.6$ and examine a redshift path $\Delta z=96$ from a 
statistical sample of 249 QSO spectra to search for $\tau_{\rm LLS} \ge 2$ LLSs.  
We compare our results with previous surveys, including the recent high redshift survey 
of \citet{prochaska10}, probing the evolution of LLSs over redshifts $0 \la z \la 5$. We connect the observational quantities to
physical properties assuming the 737 $\Lambda$CDM cosmology with $t_0=13.47$ Gyr, $H_0=70$ \kms\ Mpc$^{-1}$, $\Omega_m=0.3$, 
and $\Omega_{\Lambda}=0.7$ \citep[consistent with WMAP result,][]{komatsu09}. 

This paper is organized as follows.
After a brief description of the properties of LLSs in \S~\ref{sec:plls}, we give an overview of the data and 
the process of assembling the survey sample in  \S~\ref{sec:sample}.
In \S~\ref{sec:reds} we describe the process used to identify LLSs and characterize their properties,  
while the analysis of these properties, in particular $l(X)$ and $f(N_{\rm H I})$,  is given in \S~\ref{sec:analysis}. 
We conclude with a discussion of the connection between galaxies and LLSs in \S~\ref{sec:gal} and a summary of our principal 
results in \S~\ref{sec:sum}.

%%%%%%%%%%%%%%%%%%%%%%%%%%%%%%%%%%%%%%%%%%%%%%%%%%%%%%%%%%%%%%%%%%%%%%

\section{Description of Lyman Limit Systems}
\label{sec:plls}

The Lyman limit of neutral hydrogen is located at $\sim$912 \AA\ in the rest frame of the absorber. For a background source with intrinsic 
flux $F_{\rm QSO}$ and observed flux $F_{\rm OBS}$, the observed optical depth blueward of the limit is
\begin{equation}
\label{eqn:flux}
\tau(\lambda \leq \lambda_{\rm LLS})=\ln \frac{F_{\rm QSO}}{F_{\rm OBS}(\lambda \leq \lambda_{\rm LLS})},
\end{equation}
where $\lambda_{\rm LLS}$ is the assigned wavelength of the break in the LLS spectrum. The \HI\ column density of the absorber can then 
be related to the optical depth using  
\begin{equation}
N_{\rm H I} = \sigma_{\rm H I}^{-1}\tau_{\rm LLS}  
\end{equation}
where $\tau_{\rm LLS}$ is the optical depth at the Lyman limit of the absorption system and $\sigma_{\rm HI}=6.30~\times~10^{-18} $~cm$^{2}$ 
is the approximate absorption cross section of a hydrogen atom at the 
Lyman limit \citep{spitzer78}.

It should be noted that while we refer to the absorption systems in this survey as LLSs, a more accurate description would be 
optically thick absorbers. Since we identify all systems above a minimum $\tau_{\rm LLS}$, we limit our sensitivity 
to accurately measure large \HI\ column densities.
Strong absorbers depress the absorbed flux so low that it cannot be measured. In these cases we have only lower 
limits for the \HI\ column densities. As a result,
 some of the absorbers in the sample are likely DLAs or SLLSs, but the lack of coverage of the Lyman-$\alpha$ line 
prevents us from definitively categorizing these absorbers. Also, the frequency distribution of DLAs and SLLSs is much lower than 
for standard LLSs, suggesting the strong absorbers comprise a small portion of our sample (see \S~\ref{sec:fn}).

Due to the different selection criteria in past LLS surveys, we have created two statistical samples of our LLSs. The first sample, R1, 
defines a LLS as an absorber where $\tau_{\rm LLS} \geq 1$, i.e., $N_{\rm H I} \geq 10^{17.2}$ cm$^{-2}$. 
The majority of the surveys done through the 1990s were 
completed using this criterion, although these previous studies were not always rigorous about this restriction.
The second sample,  R2, defines a standard LLS as an absorber where $\tau_{\rm LLS} \geq 2$, i.e., $N_{\rm H I} \geq 10^{17.5}$ cm$^{-2}$. 
This second definition is adopted for comparison with the recent high redshift survey by \citet{prochaska10}. 

Although not directly included in our 
statistical analyses, we have identified many PLLSs with $\tau_{\rm LLS} < 1$, i.e., $N_{\rm H I} < 10^{17.2}$ cm$^{-2}$. 
These absorbers require a more refined assessment of their selection, and the present sample is incomplete. As a result, 
we warn against the use of such systems from our sample in statistical analyses. This incompleteness manifests 
itself in our analysis of the $f(N_{\rm H I})$ distribution for LLSs (see \S~\ref{sec:fn}).

Lastly, in dealing with QSO absorption lines, it is common to exclude absorbers located within an established distance of the 
background source to eliminate any potential influence the source may have on the number density and ionization state of the systems. 
We identify these proximate-LLSs as absorbers within $3000$ \kms\ of the background QSO and exclude them from our statistical analyses.

%%%%%%%%%%%%%%%%%%%%%%%%%%%%%%%%%%%%%%%%%%%%%%%%%%%%%%%%%%%%%%%%%%%%%%

\section{The Data: FOS and STIS}
\label{sec:sample}

In this work we make use of archival observations from both the STIS and FOS instruments on board the \emph{HST}.
The STIS  sample incorporates data from a variety of projects which used 
the G140L and G230L gratings.  These gratings are capable of a resolving power of R$\sim$1000, 
and wavelength coverages of $1150-1700$ \AA\ for the G140L and  $1600-3100$ \AA\ for the G230L. All the data were retrieved from the MAST archive and were
processed with CALSTIS v2.22 prior to retrieval. Data for objects observed more 
than once  were combined into a single spectrum weighted by the exposure 
time of the individual spectra. For objects observed with both the G140L and G230L gratings, these data 
were combined into a single spectrum.  Table~\ref{tab:stis} summarizes the observations used in this work, giving the grating  used 
for an observation, the total exposure time of the observation, and the proposal ID of the observation. 
Our final analysis of LLS statistics requires careful culling of the data
to minimize biases and some of these observations were not included in our final sample; we discuss the criteria used to exclude an observation in \S~\ref{sec:rsample}.

The FOS data can be separated into two distinct portions. First, 
we use the Bechtold et al. (2002) reductions of observations taken with the G130H, G190H, and G270H gratings.\footnote{The data are available through 
{\tt http://lithops.as.arizona.edu/$\sim$jill/QuasarSpectra/}.} 
We will refer to this subsample as FOS-H. Data taken with these 
gratings have a resolving power R$\sim$1300, and wavelength coverages of $1140-1600$ \AA\ for the G130H, $1575-2330$ \AA\ 
for the G190H, and $2220-3300$ \AA\ for the G270H.
  We also make use of FOS data using the G160L grating, and we will refer to this subsample as FOS-L. These data have a resolving power R$\sim$250 and a 
wavelength coverage of $1140-2500$ \AA. We treated these data in a manner consistent with the STIS data, with multiple exposures combined to form a single spectrum.
Table~\ref{tab:fosl} lists the observations examined for this work, giving the total exposure time of the observation as well as the proposal ID.

For a small number of objects observed with FOS, observations were taken with both the low resolution G160L grating and a combination of the 
high resolution gratings. In these cases, it is possible to detect a shift in the wavelength array of the G160L spectra relative
to the high resolution spectra. For objects where a shift was evident, the G160L spectra were shifted in wavelength space to align 
with the high resolution data. There were 20 objects where a shift in the wavelength array was detectable, of which the mean shift 
in spectrum was 4 \AA. 

The FOS spectra all suffered from background subtraction uncertainties of $\sim$30$\%$ \citep{foshb} due to the crude nature of the background determination and
lack of scattered light correction in FOS. The error vectors produced by CALFOS do not account for this background uncertainty. For regions strongly absorbed by LLSs,
the background uncertainties can dominate the error budget. To estimate this uncertainty,
we calculated  the background flux as the product of the inverse sensitivity function and the count rate for each grating. Taking $\sim$30$\%$ of this quantity
allowed us to account for the error in the initial background subtraction.

\subsection{Selection of a Statistical Sample of Absorbers}
\label{sec:rsample}

The initial sample of observations taken from the STIS and FOS archives contained $\sim$700 QSOs with redshifts $0 \la z_{\rm em} \la 3$ 
\citep[Tables \ref{tab:stis},\ref{tab:fosl};][]{bechtold02}. However, not all of these QSOs
can be used for LLS studies because the data suffer from a variety of pitfalls (i.e., poor quality or lack of coverage of 912 \AA\ and
below in the QSO rest frame) or the selection of the QSO for the STIS or FOS observation is biased in favor or against
the presence of a LLS. 

To construct a sample of QSO sightlines appropriate for studying LLS statistics we used the following approach.
We assigned the redshift of the QSO, $z_{\rm em}$, determined through emission features of the spectrum, using the results from \citet{bechtold02} when available 
and the \citet{veron10} QSO database for the remaining objects. We removed from our sample
all QSOs with no coverage below 912 \AA\ in the rest frame of the QSO or where the quality of the observation was too poor to establish an 
estimate for the continuum flux. We also excluded apparent or known broad absorption line QSOs from our sample due to the 
difficulties in studying intervening absorbers in their spectra. Next we examined the Phase II proposals for each observation to determine if any
knowledge of the sight line characteristics were known prior to the execution of the proposal that may represent a bias. 
For example, QSOs specifically targeted because \emph{International 
Ultraviolet Explorer (IUE)} data indicated the QSO was UV-bright may bias our sample against LLSs. We identified all such potentially biased observations
and removed them from our sample. There were also 2 gravitationally lensed QSOs for which we only included one of the pair in our 
sample, excluding the absorber associated with the lensing galaxy. QSOs targeted because of absorption features known from previous 
observations, such as \mgii\ absorption, DLAs, 
and 21 cm absorption represented the most common selection bias in the present sample. We did not include any LLSs in our statistical sample that 
were associated with previously identified systems toward these QSOs (these systems are listed in Table~\ref{tab:LLS} with appropriate bias indicators). 
We did however, include the redshift path covered by these QSOs and any LLSs that occurred at redshifts higher than the targeted absorber redshift. There is
a concern that including these observations, in particular the targeted strong \mgii\ absorbers, may bias our sample against detecting strong \hi\ absorbers
along the included redshift path. For the majority of the targeted \mgii\ observations, additional absorption systems along the line of sight were not accounted for 
when selecting the QSOs for observation (S. Rao, private communication, 2011).  Because of this, we believe there is no significant bias in 
including the redshift path and non-targeted LLSs of these observations. In \S~\ref{sec:analysis}
we have examined a subset of these observations to show the statistical properties of the observations are consistent with the entire statistical sample.
The remaining 249 objects listed in Table~\ref{tab:Sample} comprise our sample.

\subsection{Survey Redshift Path}
\label{sec:path}

To quantify the absorption features found in our sample, we must determine the portion of each spectral observation that is amenable to a robust search. 
This quantity is referred to as the redshift path of the survey and results from translating the observed spectral wavelengths into redshifts. 
For our survey we calculated two redshift
paths, corresponding to robust searches for LLSs defined as absorbers where $\tau_{\rm LLS} \geq 1$ and  
$\tau_{\rm LLS} \geq 2$. For these two cases, we require the local continuum flux
to exceed four times the estimated error array (i.e. $S/N \ga 4$) and to exceed two and a half times the estimated error array  (i.e. $S/N \ga 2.5$), 
respectively.  
Requiring the $S/N$ of the observation to be above this threshold allowed us to empirically define an acceptable wavelength range 
(i.e., redshift path) over which we can reliably detect LLSs. 
We also require the survey path to end at the redshift $z_{\rm prox}$ which corresponds to $3000$~\kms\ blueward of $z_{\rm em}$. 
The $S/N$ limits for our redshift path definitions were deduced through the analysis of real and simulated 
spectra; these limits correspond to our ability to detect $\tau \geq$ 1 or 2 at the 95\% confidence level.  The 
second redshift path requirement is an attempt to minimize the effect of the QSO and its environment on 
the analysis of intervening absorption systems. We note that for objects in our sample, we redefine the 
quantity $z_{\rm max}$ as the lesser value of the maximum redshift that satisfies the $S/N$ requirement and $z_{\rm prox}$.

In their recent high redshift survey ($z > 3.5$), \citet{prochaska10} note several biases that impacted their 
survey due to the presence of PLLSs.  Unidentified PLLSs in their surveys had two main effects, neither of 
which particularly impacts our survey.  In the first, unidentified PLLSs in their spectra could cause the 
local $S/N$ to drop below their threshold criterion.  These authors calculate the $S/N$ for comparison against 
their selection criterion at the wavelength of the QSO Lyman limit and use all of the available path of the 
observation, unless they are able to identify a PLLS that depresses the $S/N$ below the threshold 
at a lower redshift.  Thus, not identifying a PLLS could cause them to overestimate the redshift path 
appropriate for a given QSO.  However, we calculate a local $S/N$ at each point in our spectra and 
are able to note which regions of a spectrum are unsuitable for use in our survey.  Any effect that causes 
the $S/N$ to fall below the threshold would shorten the redshift path, even if it were not identified as, e.g., a 
PLLS.  In the second effect discussed by \citet{prochaska10}, unidentified PLLSs at redshifts just 
above a higher optical depth absorber caused them to assign a redshift for the latter too high by up to $\Delta z = 
0.1$.  This had the effect of reducing the total redshift path of their survey by a sizable amount, since their 
typical redshift path per QSO was only $\Delta z \sim0.2$.  They estimate this caused overestimates in $l(z)$ by 30 to 50\%.  This 
has a negligible effect on our survey for several reasons.  First, $\sim90$\% of the redshifts assigned in our survey 
come from measurements of Lyman-series lines associated with the LLSs rather than from the break itself.  
These measurements should be unaffected by the aforementioned bias.  Furthermore, the probability of 
having two overlapping systems, and the resulting impact on the path length calculation, are much smaller 
at the lower redshifts of our survey.  Typically our redshift path per QSO is a factor of $\sim$4 larger than the 
mean of the Prochaska survey, while the number density of absorbers is smaller by a factor of $\sim$4.  Even 
based on these considerations alone, the impact would be mitigated by more than a factor of 10.  Furthermore, because the 
number density of absorbers per unit redshift is significantly lower, the probability of having two in close 
proximity is also lower by a factor of $\sim10$.  Together these diminish the impact of this bias to  
below a $\sim$1\% effect that only impacts the sight lines without measurements based on Lyman series lines, i.e., $<10$\% of 
our sample of LLSs.  Altogether, then, these biases play little role in our survey.  

 Table~\ref{tab:Sample} summarizes the properties of the QSO sight lines that meet these selection criteria. For each object, we give the emission redshift,
 $z_{\rm em}$, and the maximum and minimum redshifts meeting our redshift path criteria for each optical depth regime, $z_{\rm max}$ and  $z_{\rm min}$, where
$z_{\rm min}$ corresponds to the greater value of the minimum redshift that satisfies the $S/N$ requirement and 20 \AA\ above the minimum wavelength coverage of 
the observation. 
We refer to the QSO sightlines in which we can 
reliably detect a LLS with $\tau_{\rm LLS} \geq 1$ as the R1 sample, which contains $229$ QSOs and $61$ LLSs, while the objects in 
which we can reliably detect a LLS with $\tau_{\rm LLS} \geq 2$ is the R2 sample, which contains $249$ QSOs and $50$ LLSs.

\begin{figure}
\epsscale{1.0}
\plotone{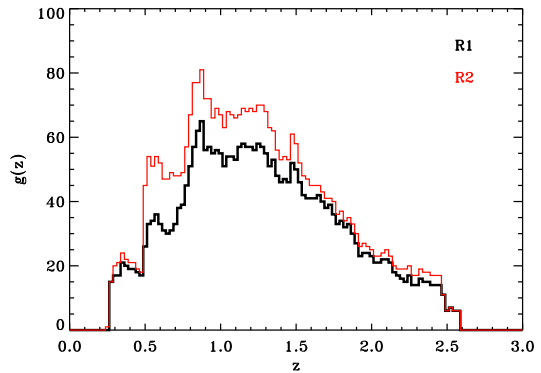}
\caption{The redshift path surveyed with our samples of $\tau_{\rm LLS} \ge 1$ (black) and $\tau_{\rm LLS} \ge 2$ (red) spectra. The function $g(z)$ is the 
number of unique QSOs in our \textit{HST} sample that probe the redshift interval $\Delta z= 0.025$, at redshift $z$, for LLSs. }
\label{fig:g_z}
\end{figure}

Figure~\ref{fig:g_z} shows the $g(z)$ distributions, which represents the number of QSOs with spectral 
coverage of $\lambda_{\rm LLS}$, as a function of redshift for the R1 and R2 samples. Both samples are most sensitive to the 
detection of LLSs over the redshifts $0.8 \le z \le 1.5$. The total integrated redshift path,
\begin{equation}
\label{eqn:gz}
\Delta z = \int g(z) dz
\end{equation}
is  $\Delta z(R1) = 79$ and  $\Delta z(R2) = 96$. Our survey probes a factor of $\ga 4$ larger redshift path than previous surveys at $z < 2.6$
\citep[$\Delta z = 21$,][]{jannuzi98}. Our survey probes a redshift path very similar to the recent high redshift survey of \citet{prochaska10}, where
$\Delta z = 96$ for LLSs at $3.3 \le z \le 5.0$.

%%%%%%%%%%%%%%%%%%%%%%%%%%%%%%%%%%%%%%%%%%%%%%%%%%%%%%%%%%%%%%%%%%%%%%%

\section{Identifying and Characterizing Lyman limit Systems}
\label{sec:reds}

\begin{figure}
 \epsscale{1.0}
 \plotone{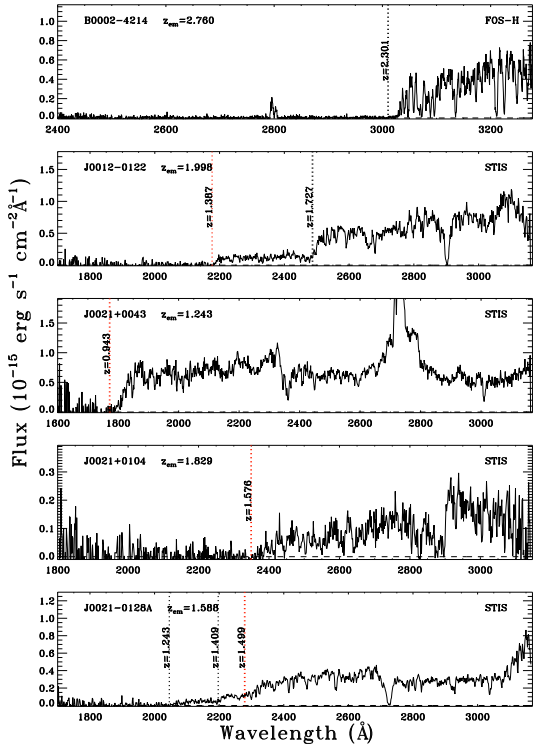}
 \caption{The first five observations of the LLSs listed in Table~\ref{tab:LLS}. The vertical, dashed lines represent LLSs included in sample R1/R2.
The vertical, dashed red lines represent the absorbers that were identified but not included in the statistical analysis for various reasons (see Table~\ref{tab:LLS} for more information). 
Spectra for each LLS listed in Table~\ref{tab:LLS} can be found in the online version.}
 \label{fig:lls}
 \end{figure}

We select LLSs on the basis of their Lyman limit absorption  (i.e., we do not include absorbers in our statistical sample based
only on strong line absorption) for redshifts where the data satisfied our redshift path criteria. The entire list of 206 LLSs found while examining our unabridged 
sample is given in Table~\ref{tab:LLS}. The absorbers used in the statistical analysis are designated with R1 or R2. There are 61 LLSs in the R1 sample and
50 LLSs in the R2 sample. 
A sample of the spectra for the observations can be found in Figure~\ref{fig:lls}, where each QSO spectrum is plotted with 
a vertical dashed line at the location of an established \HI\ absorber. The red dashed lines indicate systems which were identified but not included in our statistical analysis.
The complete sample of LLSs identified in this work are available in the online version of Figure~\ref{fig:lls}.
 
In general, as seen in Figure~\ref{fig:lls}, the break produced 
by a $\tau_{\rm LLS} \geq 1$ LLS is abrupt enough to be found in even low S/N ($\sim4$) and resolution spectra. 
However, as we discussed above the occasional presence of PLLSs can complicate the situation. In particular, assigning the continuum flux level redward 
of the Lyman break can become difficult. To minimize the potential error associated with this effect, we adopt a two-step process. First 
 we use an automated search to identify potential Lyman limits. This automated search was checked by-eye and found to highlight absorbers
with $\tau \ge 1$ quite well. These methods allowed us to identify absorbers
where $\tau < 1$, but we stress the sample of PLLSs detected is not complete. Subsequently we use an interactive routine to fit the continuum flux, 
 the optical depth of the system, and the characteristic continuum recovery blueward of a Lyman limit (see below). While our 
statistical sample contains only LLSs that satisfy our $\tau_{\rm LLS} \geq 1$ or $\tau_{\rm LLS} \geq 2$ criteria, we have attempted to identify 
every optically thick and partially optically thick absorber present in our spectra. This is important for accurate continuum fitting and 
provides a sample of PLLSs that we use in the analysis of the $f(N_{\rm H I})$ distribution presented in \S~\ref{sec:fn}. 

We adopted the composite QSO spectrum developed by \citet{zheng97} as a general model of the QSO continuum.
We scaled the continuum to each QSO spectrum over a relatively absorption free wavelength range of the spectrum. We found the majority
of QSO observations were fitted well by this composite. We then used a running chi-square tool to identify portions
of the spectrum where the QSO spectrum deviated from the composite. For each pixel in the spectrum, a running $\chi^2$ goodness of fit parameter was 
calculated comparing the fitted continuum with the observed spectrum over $\sim$30 \AA.\footnote{At low redshift the attenuation of the QSO flux
blueward of 912 \AA\ (in the rest frame of the QSO) due to intervening Lyman-$\alpha$ lines is quite small compared to high redshift.
 This allowed us to model the QSO flux with the composite spectrum quite well over all wavelengths, including regions which probed the lower redshift
Lyman-$\alpha$ forest.} 
 This largely excluded false identifications due to 
strong absorption lines present throughout the spectrum. Spectra not well fitted by this routine were individually 
examined for the possibility of a Lyman break, although the number of such spectra is very small.

\begin{figure*}
 \epsscale{1.0}
 \plotone{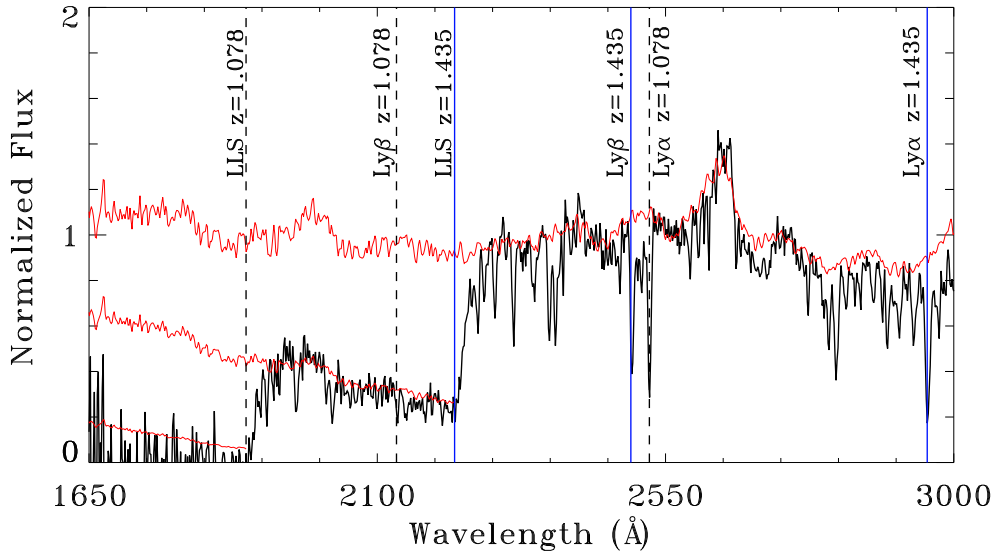}
 \caption{The spectrum of J1322+4739 with a composite spectrum overplotted (in red). The Lyman limit at $z=1.435$ $(\tau = 1.27 \pm 0.04)$ is identified with 
 blue lines. Blueward of the limit, 
the composite is overplotted again with the characteristic recovery of the flux. The Lyman limit at $z=1.078$ $(\tau > 2.44)$ is identified with the 
dashed black lines. This is a snapshot of the plots produced 
using our method to identify LLSs and characterize their properties.}
 \label{fig:method}
 \end{figure*}

Once a spectrum was flagged as containing a possible LLS, the spectrum was examined more thoroughly to identify the redshift of the 
break and any Lyman series lines present. When possible, the Lyman series lines were used to determine the redshift of the absorber. However, if the Lyman limit was located near the 
maximum wavelength of the spectrum or the resolution of a spectrum was too low to identify individual absorption lines 
(i.e. the majority of FOS-L observations) we used the Lyman break to set the redshift. We define the redshift of a LLS determined from the break as
\begin{equation}
z_{\rm LLS}=\frac{\lambda_{\rm LLS}}{912 {\rm \AA}}-1,
\end{equation}
where $\lambda_{\rm LLS}$ is the wavelength of the observed Lyman break. In Table~\ref{tab:LLS}, we 
list  $z_{\rm LLS}$ from our analysis. The typical statistical error on the redshift determination from the Lyman series lines is $\sigma_z=0.001$ 
for the FOS-H and STIS spectra. The statistical errors on $z_{\rm LLS}$ are larger when using the break, about 0.010 for the FOS-H and STIS spectra 
and  0.014 for the FOS-L spectra. As we used two different methods to determine $z_{\rm LLS}$, possible systematics may be introduced. 
We tested this by using LLSs for which the redshift could be determined from both the Lyman lines and break. We found a systematic shift 
of 0.007 in the redshifts determined from the break in the FOS-H and STIS spectra, but not for the FOS-L (possibly because the resolution 
is far cruder). For the few redshifts of the LLS determined from the Lyman break in the FOS-H and STIS spectra, we systematically corrected 
$z_{\rm LLS}$ by the 0.007 shift.

Once a redshift is assigned, we measure the optical depth at the Lyman limit for each absorber. We interactively refined the fit of the 
composite QSO continuum model to each spectrum. We determine the optical depth at the limit by comparing the continuum flux, 
$F_{\rm QSO}$,  with the observed (absorbed) flux, $F_{\rm OBS}$, as in Equation~\ref{eqn:flux}. For many PLLSs and a few LLSs, the residual flux below the limit is sufficient
to satisfy our $S/N$ selection criteria for further LLS searches at $z < z_{\rm LLS}$ of the highest redshift system. 
We derive a continuum blueward of the highest redshift LLS in a spectrum by modeling
the recovery of the flux due to the wavelength dependence of the optical depth. The continuum flux in the recovery region, $F_{\rm REC}$, is
\begin{equation}
\label{eq:frec}
F_{\rm REC} =  F_{\rm QSO} e^{-\tau_{\rm LLS} \left( \frac{\lambda }{912 {\rm \AA}} \right)^3} , {\rm for} \lambda < 912 {\rm \AA}.
\end{equation}

Once $F_{\rm REC}$ is defined, we repeat our LLS search for systems at redshifts below the initial system after renormalizing 
our best fit continuum fit according to Equation~\ref{eq:frec}. Figure~\ref{fig:method}  
shows the method of fitting the continuum onto a QSO spectrum, identifying a LLS, and modeling the recovery of the 
spectrum blueward of a Lyman limit. 

For absorption systems where the residual absorbed flux blueward of the break was determined to the $3\sigma$ level, we report optical depth measurements with accompanying errors.
For systems where we could not detect the residual absorbed flux to the $3\sigma$ level, we treat the optical depth measurement as a lower limit and report the $2\sigma$ lower limit.
The optical depth measurements can be found in Table~\ref{tab:LLS}, where $N_{\rm HI}$ is also reported.

%%%%%%%%%%%%%%%%%%%%%%%%%%%%%%%%%%%%%%%%%%%%%%%%%%%%%%%%%%%%%%%%%%%%%%%

\section{Statistical Analysis and Results}
\label{sec:analysis}

In this section we present the results of our survey. The first subsection examines the redshift density of LLSs and how the results from 
sample R1 and R2 compare to past studies of LLSs. The following subsections introduce a $\Lambda$CDM cosmology to connect the statistical 
treatment of our samples to physical structures throughout the universe such as the mean separation of LLSs, the incidence of LLSs
as a function of absorption distance, and the column density distribution function. In each subsection, we first generalize the analysis, 
as to make it applicable to both of our samples. Following the general treatment, the individual samples are explored and discussed
when appropriate. 

\subsection{The Redshift Density of Intervening LLSs}

\begin{figure}[!h]
 \epsscale{1.1}
 \plotone{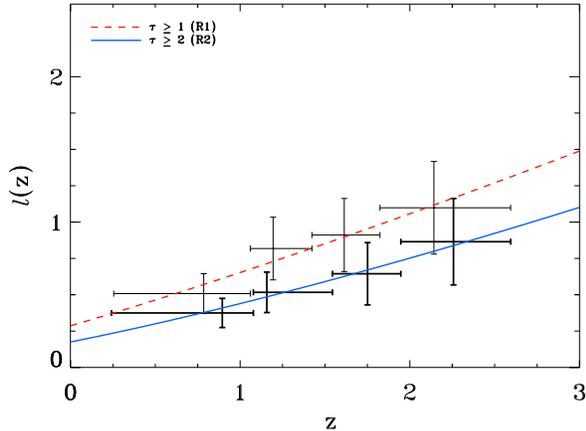}
 \caption{The evolution of the redshift density, $l(z)$. The values for $l(z)$ can be found in Table~\ref{tab:stats}. 
 From a maximum-likelihood analysis, the best fit power law: $l(z)=l_*[(1+z)/(1+z_*)]^\gamma$, with 
$z_*=1.5$, is $l_*=$ \None\ and $\gamma=$ \gammaone\ (for $\tau_{\rm LLS} \geq 1$) and $l_*=$ \Ntwo\ and $\gamma=$ \gammatwo\ (for $\tau_{\rm LLS} \geq 2$).}
 \label{fig:dndz}
 \end{figure}

The redshift density of LLSs, $l(z)$,\footnote{This quantity has often been denoted in the past by a variety of symbols 
including $N(z)$, $n(z)$, and $dN/dz$.} is a statistical quantity that is directly related to the QSO observations.  
The standard method for estimating $l(z)$ is to simply calculate the ratio of the number of LLSs, $N$, detected in a redshift interval 
to the total survey path, $\Delta z$ (defined in Equation~\ref{eqn:gz}), contained in that redshift interval:
\begin{equation}
l(z)=\frac{N}{\Delta z}.
\end{equation}
Figure~\ref{fig:dndz} presents the values of $l(z)$ for both samples, R1 and R2. We first estimated $l(z)$ in redshift intervals where the binning was arbitrarily selected to 
provide approximately the same number of LLSs in each interval. Table~\ref{tab:stats} lists the properties of these redshift intervals
for the R1 and R2 samples. Following previous work \citep[e.g.][]{tytler82}, we model the redshift evolution in $l(z)$ as a power law of the form:
\begin{equation}
\label{eqn:nz}
l(z)=l_*\left(\frac{1+z}{1+z_*}\right)^\gamma.
\end{equation}
This functional form was originally chosen when the Einstein-de Sitter models were the preferred cosmologies. At the time, 
evolution in the LLS distribution was found if $\gamma \neq 1$ for $q_0=0$ or $\gamma \neq 0.5$ for $q_0=0.5$. We use this functional
form for the historical significance and the usefulness it provides in comparing our results with previous surveys, but we note there is no physical 
or a priori reason to expect a particular functional form. However, the power law fit does do a reasonable job fitting the $l(z)$ distribution.

Using the maximum-likelihood method \citep[e.g.][] {tytler82,sargent89}, a best-fit estimate for $\gamma$, 
and from that $l_*$, can be determined for both samples. For the R1 sample ($\tau_{\rm LLS} \geq 1$) we find
 $\gamma =$~\gammaone\ and $l_*=$~\None. For sample R2 ($\tau_{\rm LLS} \geq 2$) we find $\gamma=$~\gammatwo\ and $l_*=$~\Ntwo. For both samples we adopt $z_*=1.5$, which corresponds to $\langle z_{LLS} \rangle=1.5$ and can be chosen arbitrarily.
 These best-fit models are overplotted on the $l(z)$ data in Figure~\ref{fig:dndz}. 

To check if the difference between the observed distribution and the adopted power law expression is statistically significant, we test the 
null hypothesis, that the observed and predicted cumulative distributions of LLSs with redshift are distinct, using a Kolmogorov-Smirnov test.  
The KS test yields a minimum probability of P=0.95 that we can reject the null hypothesis, using the entire redshift range encompassed by 
both the low and high redshift samples.  Thus there is a strong probability that we can reject this null hypothesis.

 As mentioned in \S~\ref{sec:rsample}, we examined the potential biases associated with including redshift paths toward QSOs with targeted strong \mgii\ absorbers,
which constitute a significant
fraction of our statistical sample ($\sim 25$\%). To empirically test for any bias, we separately analyzed the statistical properties of these observations and compared
their properties  with the statistical properties of the total sample. From the STIS observations of Rao et al. (PID 9382 \& 8569), we composed a sample of 79 QSO 
observations. This sample contained 17 (16) $\tau \geq 1$ (2) LLSs over a redshift path of $\Delta z = 22.65$ (28.01), giving $l(z) = 0.75\pm0.20$ ($0.57\pm0.14$). 
These values are well within $1\sigma$ of the $l(z)$ for the full sample ($0.77\pm0.11$ and $0.52\pm0.08$ for $\tau \geq 1$ and 2, respectively; Table~\ref{tab:stats}). 
As a further check, we then separately analyzed the remaining 170 QSO observations to compare with the statistical properties of the total sample. This sample contained
44 (34) $\tau \geq 1$ (2) LLSs over a redshift path of $\Delta z = 56.57$ (68.13), giving $l(z) = 0.78\pm0.13$ ($0.49\pm0.09$), which again is well within the $1\sigma$ 
values for the full sample (Table~\ref{tab:stats}). This suggests any biases associated with these observations have a negligible impact on our analysis and results.

\begin{figure}[!h]
 \epsscale{1.1}
 \plotone{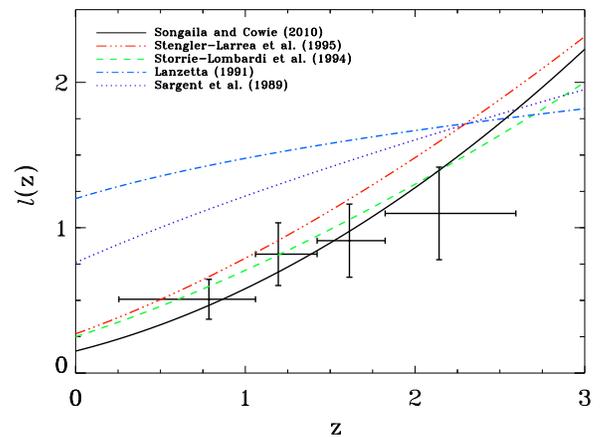}
 \caption{Estimates for the functional form of $l(z)$ from previous studies of $\tau_{\rm LLS} \geq 1$ LLSs plotted on top of our R1 sample. The fits are parameterized as power laws with individual 
parameters listed in Table~\ref{tab:comp_stats}. Our results are consistent with the results of Stengler-Larrea et al. (1995) 
and Storrie-Lombardi et al. (1994), studies which probed the range $0 < z < 4$.}
 \label{fig:compare}
 \end{figure}

Over the past 30 years, there have been a variety of LLS surveys \citep{tytler82,sargent89,lanzetta91,storrie94,stengler95,jannuzi98,prochaska10,
sc10},
 and, as a result, a variety of estimates of $l(z)$. 
Many of these previous surveys examined large redshift intervals (typically spanning $0 \la z \la 4$) but have largely been statistically  
dominated by high redshift ($z \ga 2.5$) LLSs. It is due to this inhomegeneity, combined with the lack of low redshift LLSs, that there is uncertainty as to the 
true statistical distribution of LLSs over the redshift range $0 \la z \la 4$. \citet{lanzetta91} was the first to argue for a potential break in the 
evolution of the redshift density of LLSs at $z \sim 2.5$, where he found the low redshift ($z < 2.5$) LLSs showed relatively constant $l(z)$ and the high redshift ($z > 2.5$) 
LLSs showed a rapidly evolving $l(z)$. However, both \citet{storrie94} and \citet{stengler95} argued, based on
samples spanning the redshift range $0 \la z \la 4$, that the $l(z)$ for LLSs is best described as moderately evolving over the entire range and 
fit by a single power law in $(1+z)$.
Figure~\ref{fig:compare} presents $l(z)$ values from the R1 sample, as well as the $l(z)$ fits from these previous surveys (with the parameters listed in Table~\ref{tab:comp_stats}). 
These previous surveys used a $\tau_{\rm LLS} \geq 1$ criterion for inclusion in the sample, but it is not clear this was applied in a uniform manner \citep{stengler95}.
Our results for $l(z)$ over $0.25 \leq z \leq 2.6$ are consistent with the surveys of \citet{storrie94} and \citet{stengler95}, 
both of which fit $l(z)$ over the redshift range $0 \la z \la 4$.

Recently, \citet{prochaska10} released a survey of high redshift LLSs (with $\tau_{\rm LLS} \geq 2$) using the SDSS-DR7 that samples a redshift
range  ($3.5 \leq z \leq 4.4$) that does not overlap our survey redshifts. They find the $l(z)$ for high redshift LLSs can be described as rapidly evolving over the range $3.5 \leq z \leq 4.4$. 
Our model of $l(z)$ is inconsistent with the \citet{prochaska10} survey when extrapolated to $z > 3$, as the Prochaska results are inconsistent with ours
if extrapolated to $z<2.6$. It was a similar disagreement seen in the high and 
low redshift samples of the \citet{lanzetta91} work that led to the argument for a break in the power law description of $l(z)$ for LLSs. To investigate
the possibility and significance of a broken power law fit to the redshift density of LLSs, we combine the recent high redshift sample 
from \citet{prochaska10} with our low redshift sample to examine the statistical nature of LLSs from $0 \la z \la 5$.%\footnote{J.X. Prochaska kindly made their data available 
%for comparison with this sample prior to publication.}

\begin{figure}[!h]
 \epsscale{1.2}
 \plotone{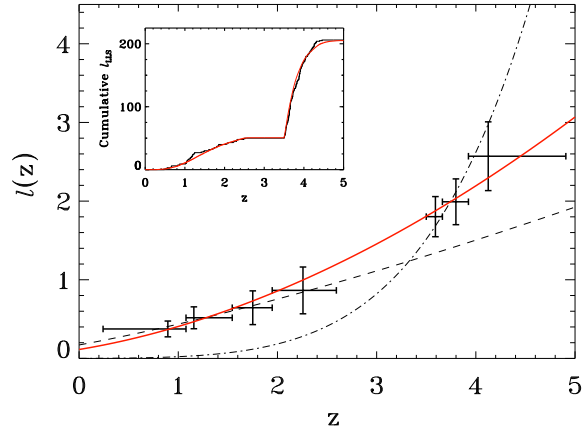}
 \caption{  The incidence of LLSs per redshift, $l(z)$, for the combined SDSS-DR7 and R2 samples (RP10, $\tau \ge 2$). The single power law:  $l(z)=l_*[(1+z)/(1+z_*)]^\gamma$, 
with $z_*=3.23$, is $l_*=$ \Nc\ and $\gamma=$ \gammac, is plotted with the $l(z)$ values taken from Table~\ref{tab:stats2}. To check the statistical significance of the single power law, a KS test was 
administered to the sample. The cumulative distribution function for the observations overplotted with the predicted distribution from 
our best fit power law can be found in the insert. The KS test gives a probability $P=0.95$ that we can reject the null-hypothesis 
that the two distributions  originate from different populations. The dashed and dot-dashed curves are the power law fits derived for the low and high redshift samples respectively.}
 \label{fig:re_evolve}
 \end{figure}

 We refer to the combined  R2 and \citet{prochaska10} samples as the RP10 sample.  
This combined sample of \emph{HST} and SDSS observations contains 685 QSOs and 206 LLSs with $\tau_{\rm LLS} \geq 2$. 
The total redshift path probed in RP10 is $\Delta z = 172$. This redshift path is a factor of $\sim$2 greater than the recent LLS survey from
\citet{sc10}, which spanned redshifts up to $z\sim6$. In Figure~\ref{fig:re_evolve}, 
we present our estimate for $l(z)$ over this expanded redshift range.  We find $l(z)$ from the combined sample can be described by a single power law 
(Equation~\ref{eqn:nz}) with $\gamma =$~\gammac\ and $l_*=$~\Nc, for $z_*=3.23$. Table~\ref{tab:stats2} lists the properties of the $l(z)$ 
bins used for display purposes in Figure~\ref{fig:re_evolve} and the values associated with each bin. To confirm the observations are well 
modeled by a 
 single power law, a KS test was applied to the cumulative distribution function of observed and predicted LLSs(see 
Figure~\ref{fig:re_evolve}). The KS test yields a probability of at least $P=0.95$ that the null hypothesis, the observed and predicted distributions represent 
different distributions, can be discarded. Thus the RP10 sample supports the conclusions of \citet{storrie94}, \citet{stengler95}, and more recently \citet{sc10},
that a single power law is sufficient
to describe $l(z)$ over $0<z<5$.

It should be noted that in the original analysis of the SDSS-DR7 sample, \citet{prochaska10} limited the redshift path to $z \leq 4.4$. This was
done because the inclusion of $z > 4.4$ into their sample appeared to produce an artificially low $l(z)$ result for $4.4 < z < 5.0$, which they argued 
was unlikely to be physical. We include this extra redshift path from their sample, under the reasoning that arbitrary binning of the data for display purposes can
produce artificial departures from a trend that in no way affects the statistical analysis of the maximum likelihood method. In our redefined bins, the artificial 
drop apparent at $z > 4.4$ is no longer present.  
To insure the extra redshift path is not solely responsible for 
our ability to fit the combined sample with a single power law, we conducted our analysis on the combined \emph{HST}/SDSS for 
both situations ($z_{\rm max} \leq 4.4$ and unrestricted $z_{\rm max}$ ) and found both conditions produce single power law fits that 
are consistent and good describers of the data. It is also interesting to note that if $z_{\rm max}$ is unrestricted for the SDSS sample alone, the best fit curve for the 
high redshift sample is described by $\gamma=2.79 \pm 1.46$ and $l_*=1.94$. This description of $l(z)$ for the high redshift LLSs  
presents a less convincing argument for the need of a break in the power law, because the difference in power law indices between low
and high redshift is less extreme.  

Our analysis indicates that it is not necessary to introduce a broken power law to model the statistical evolution of the redshift 
density of LLSs over $0 \la z \la 5$.  However, we stress a sample with coverage of the $2.5 \la z \la 3.5$  region will be needed to truly rule
out a break \citep{omeara10}. We note that the redshift density is
really an observational statistic, and the difference between a single or broken power law may not carry much significance over to the
physical quantities with which it is related. In \S~\ref{sec:dndx}, \S~\ref{sec:dr}, and \S~\ref{sec:fn}
 we put these results into the context of a cosmology and discuss the implications for the evolution and nature of LLSs to $z\sim$5.

\subsection{ The Incidence of LLSs per Absorption Distance}
\label{sec:dndx}
 The number of LLSs per absorption length, $l(X)$ \citep{bp69}, is defined as
\begin{equation}
\label{eqn:nx}
l(X)dX=l(z)dz
\end{equation}
where
\begin{equation}
\label{eqn:dx}
dX=\frac{H_0}{H(z)}(1+z)^2dz,
\end{equation}
and
\begin{equation}
\label{eqn:hz}
 H(z)=H_0(\Omega_{\Lambda}+\Omega_{m}(1+z)^3)^{1/2}.
\end{equation}
 The quantity $l(X)$ is defined  such that it is constant if the product of the comoving number density of structures giving rise to LLSs, $n_{\rm LLS}$,
and the average physical cross section of the structure, $\sigma_{\rm LLS}$, is constant, i.e., $l(X) \varpropto n_{\rm LLS}\sigma_{\rm LLS}$.

\begin{figure}[!h]
 \epsscale{1.15}
 \plotone{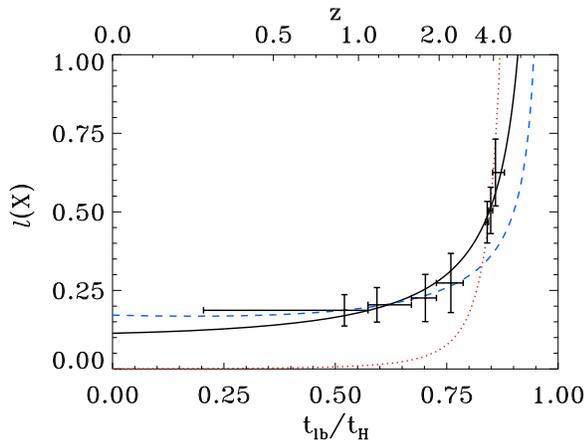}
 \caption{ The incidence of LLSs per absorption distance, $l(X)$, plotted as a function of fractional look-back time. The values for $l(X)$ can be found in Table~\ref{tab:stats2}.
The solid line is the power law result for the maximum-likelihood
analysis on the RP10 sample ($\tau \ge 2$). The dashed blue line and the dotted red line are the power law fits from the R2 sample and high redshift
sample from \citep{prochaska10}. This figure illustrates the problems that arise when attempting to fit the high and low redshift regimes with power law fits derived in either the high or low redshift
only. }
 \label{fig:look}
 \end{figure}

Figure~\ref{fig:look} shows the quantity $l(X)$ plotted as a function of fractional lookback time for the RP10 sample ($\tau_{\rm LLS} \geq 2$). We see that  
$l(X)$ experiences a rapid decrease for $\sim$0.6 Gyr corresponding to a decrease in redshift from $z=4.9$ to 3.5.  After this rapid drop, $l(X)$ decreased slowly over 
 $\sim$8 Gyr, from $z=2.6$ to $0.25$ (See Table~\ref{tab:stats2}). The results in Table~\ref{tab:stats2} show that $l(X)$ fell by a factor of 
$\sim$1.5 over $\sim$0.6 Gyr at high redshift and by another factor of 1.5 over $\sim$8 Gyr at low redshift.

Figure~\ref{fig:look} demonstrates why differing results are found regarding the broken (or not) power laws in the statistical treatment of the redshift density of LLSs.
 The dashed red line  and dotted blue line in the plot are the best fit power laws for 
$l(z)$ (transformed into $l(X)$ using Equation~\ref{eqn:nx} and \ref{eqn:dx}) for our low redshift sample and the high redshift sample of \citet{prochaska10}. The solid black line
is the best fit power law to the RP10 sample (again transformed into $l(X)$ using Equation~\ref{eqn:nx} and \ref{eqn:dx}). The nature of power laws makes it difficult to extrapolate
a fit based on observations in only the low or high redshift regime (in the regime where the power law is derived, the fit is nearly linear, making it extremely difficult to match
observations in a regime outside of where it was derived). It is only when  the  observations are combined that we are able to produce
a consistent single power law.  We have mentioned 
the need for a study of the intermediate redshift regime \citep[$2.5 \leq z \leq 3.5$,][]{omeara10},  which will allow for a more definitive assessment of the absorber distribution.

As previously stated, the behavior in $l(X)$ is related to the comoving number density of LLSs as well as the physical size of the absorbers. This rapid decrease
in $l(X)$ over a short timescale at high redshift indicates either the physical size of LLSs has decreased substantially in this time or the comoving number density of LLSs has dropped
significantly. A moderate decrease in both properties could also give rise to this behavior, but as we will show in \S~\ref{sec:gal}, when we associate LLSs with galaxies we find the physical 
size of LLSs must undergo significant evolution from $z\sim$5 to 2.  

\subsection{ The Mean Proper Separation of LLSs}
\label{sec:dr}

The number density of optically thick absorbers throughout the Universe determines the mean free path of hydrogen ionizing photons, and in turn, sets the shape and intensity of the UVB. 
We can calculate an upper limit to this mfp using $l(X)$ of $\tau_{\rm LLS} \geq 2$ absorbers, as calculated in \S~\ref{sec:dndx}. It is only an upper limit
because we have not included the $\tau_{\rm LLS} < 2$ absorbers that contribute to the overall absorption of hydrogen ionizing photons. Using $l(X)$, we can calculate the average proper distance, 
$\Delta r_{\rm LLS}$, a photon travels before encountering a $\tau_{\rm LLS} \geq 2$ LLS \citep[e.g., ][]{prochaska10} as
\begin{equation}
\Delta r_{\rm LLS} = \frac{c}{H_0}\frac{1}{(1+z)^3 l(X)}.
\end{equation}

\begin{figure}[!h]
 \epsscale{1.15}
 \plotone{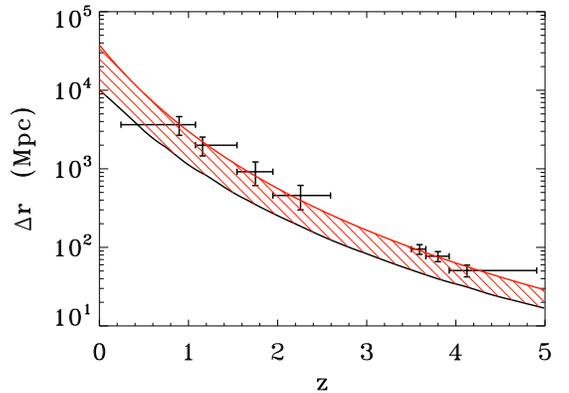}
 \caption{The average proper distance a photon travels before encountering a $\tau \geq 2$ LLS, plotted as a function of redshift. The data points correspond to values taken from
Table~\ref{tab:stats2}, with the solid red line representing the interpolated functional form for $\Delta r_{\rm LLS}$. We have included in this plot 
the mean free path of a hydrogen ionizing photon (solid black line), taken from the recent work on the ionizing background spectrum by 
Faucher-Giguere et al. (2009). The shaded region highlights the difference in the two curves, which corresponds to the effect PLLSs and $\tau_{\rm LLS} < 2$ LLSs 
(which were not included in our RP10 statistical sample) have on the opacity of the universe. }
 \label{fig:meanr}
 \end{figure}

With the RP10 sample we find that $\Delta r_{\rm LLS}$ varies from $\sim 50 - 3700$ $h_{70}^{-1}$ Mpc proper distance from $z\sim$5 to 0.3  
(Table~\ref{tab:stats2}, also note $\Delta r_{\rm LLS}$ was calculated for R1 and R2 in Table~\ref{tab:stats}).  
In Figure~\ref{fig:meanr}  $\Delta r_{\rm LLS}$ is shown as a function of redshift (data points 
and red curve), along with the mean free path of hydrogen ionizing photons (black curve) estimated by \citet{faucher09} (It should be noted that while the
calculation of the mean free path by \citet{faucher09} is dependent on an assumed \HI\ distribution, their estimated mean free path is in agreement with the
mean free path calculation from \citet{prochaska09}.). 
The shaded region emphasizes the difference between the two curves, which
can be associated with the contribution from PLLSs and $\tau_{\rm LLS} < 2$ LLSs that were not included in this calculation. We note the ratio 
between the distance a photon travels before encountering a $\tau_{\rm LLS} \geq 2$ LLS and the mean free path of a hydrogen ionizing photon is increasing with decreasing redshift. Consider the extreme redshifts
of the plot;  at $z\sim$5, $\Delta r_{\rm LLS}$ is a factor of $\sim$1.5 larger than the predicted mean free path, while at $z\sim$0, $\Delta r_{\rm LLS}$ is a factor of $\sim$3.5 larger than the predicted mean free path.
Assuming a mean free path consistent with \citet{faucher09}, this suggests that the 
$\tau_{\rm LLS} < 2$ hydrogen absorption systems have become increasingly more important for absorption of Lyman continuum photons as the universe has evolved.  

\subsection{The Differential Column Density Distribution Function}
\label{sec:fn}

In this subsection, we combine our low redshift sample with previous works on the low-$z$ ($z \la 2.6$) IGM to place constraints on the differential column density distribution $f(N_{\rm H I})$ over 10 orders of magnitude in $N_{\rm HI}$. This distribution is defined such that $f(N_{\rm H I},X) dXdN_{\rm H I}$ is the number of absorption systems with column density between $N_{\rm H I}$ and $N_{\rm H I}+dN_{\rm H I}$ and redshift path between $X$ and $X+dX$ 
\citep[e.g.,][]{tytler87}, 
\begin{equation}\label{e-fn1}
f(N_{\rm H I}) dN_{\rm H I} dX =  \frac{m}{\Delta N_{\rm H I} \Sigma \Delta X  } dN_{\rm H I}dX\,,
\end{equation}
where $m$ is the observed number of absorption systems in a column density range $\Delta N_{\rm H I}$ centered on $N_{\rm H I}$ and $ \Sigma \Delta X  $ is the total absorption distance covered by the spectra. The first moment of the distribution is also the incidence of absorbers per absorption distance, $l(X) = \int_{N_1}^{N_2} f(N) dN$. Empirically, it has been shown that at low and high redshift, $f(N_{\rm H I})$ may be  fitted by a power law for various $N_{\rm H I}$ regimes \citep[e.g.,][]{tytler87,rao06,lehner07,omeara07}: 
\begin{equation}\label{e-fn2}
f(N_{\rm H I})dN_{\rm H I}dX = C_{\rm H I} N_{\rm H I}^{-\beta}dN_{\rm H I}dX\,.
\end{equation}

\begin{figure*}
 \epsscale{1.0}
 \plotone{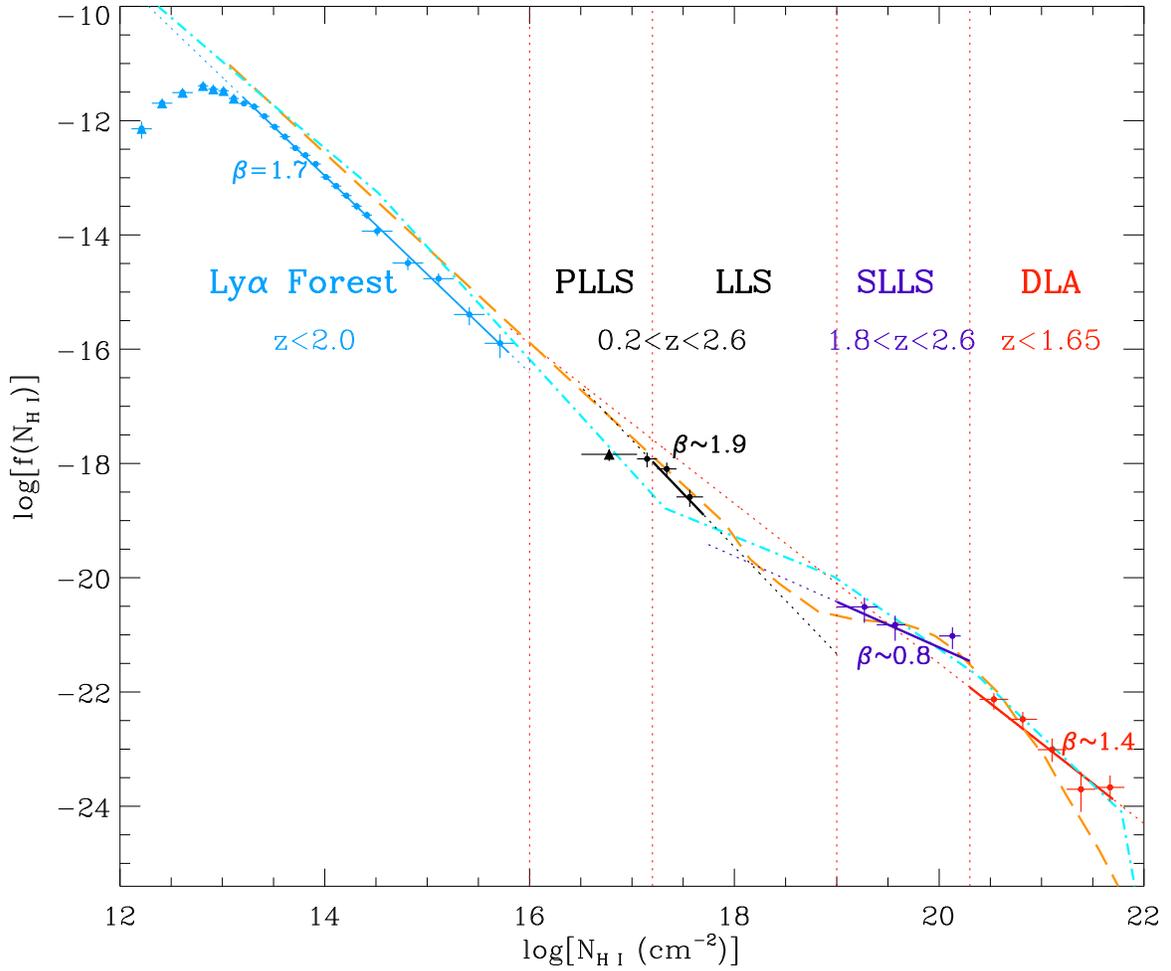}
 \caption{The differential density distribution ($f(N_{\rm H I})$) is plotted against $\log N_{\rm H I}$. The data are shown with filled circles or triangles and error bars.  The triangles indicate that the sample is incomplete at these column densities. The blue, black, violet, and red solid curves are maximum-likelihood fits to the Lyman-$\alpha$ forest, LLS, SLLS, and DLA samples. The solid part of these lines shows the $N_{\rm H I}$ portion where the fit was undertaken, and the dotted part is the extension in other $N_{\rm H I}$ regimes. The dot-dashed cyan curve is the estimation of $f(N_{\rm H I})$ at $z\sim3.7$ by \citet[][the solid black curve in their Figure~14; bear in mind that for $14.5 \le \log N_{\rm H I}< 19$, $f(N_{\rm H I})$ is quite uncertain]{prochaska10}. The orange long-dashed curve is a model from \citet{corbelli02} at low $z$ where $f(N_{\rm H})$ is assumed to follow a single power law, but $f(N_{\rm H I})$ deviates from a single power law as a result of photoionization by the UV background (see \S\ref{sec:fn} for more details).}
 \label{fig:fn}
 \end{figure*}

The slope, $\beta$, may vary with the considered $z$ or $N_{\rm H I}$ intervals, and, as discussed below and elsewhere \citep[e.g.,][]{wolfe05,pw09}, the functional form can be more complicated than a single power law, especially when the entire observed $N_{\rm H I}$ range is considered.  In Figure~\ref{fig:fn}, we show the $f(N_{\rm H I})$ column density distribution at $z\la 2.6$. The data and analyses for different $N_{\rm HI}$ regimes come from various origins that we detail below. In the studies where another cosmology was chosen to calculate $\Delta X$ \citep{lehner07,williger10}, we have updated the cosmology  to that used in the present study (see Equation~\ref{eqn:dx}). At $z<1.65$, the DLA sample was selected based on known strong \mgii--\feii\ systems \citep{rao06}.  Their sample consist principally of data similar to those presented in this work (but rejected from our sample of LLSs because they were specifically targeted) with the addition of {\it IUE} spectra. Owing to their selection criteria, the sample has selection biases \citep{rao06,pw09}, although \citet{rao06} argued that they are relatively well understood and dealt with (in the DLA regime). \citet{rao06} found that their DLA ($\log N_{\rm H I} \ge 20.3$) sample could be fitted with $\beta = 1.4$ (represented by the solid red line in Figure~\ref{fig:fn}). The dot-dashed cyan curve shows a fit assuming the power law index for the DLA at high redshift with $\beta = 1.8$ ( $20.3 \la \log N_{\rm H I}\la 21.8$) \citep{prochaska10}, which seems to provide a reasonable fit to the DLA measurements for $z<1.65$ as well. The similar slope of $f(N_{\rm H I})$ at both high and low $z$  is consistent with a non-evolution of $f(N_{\rm H I})$ for the DLA as argued by \citet{pw09}. 

At the other end of the $N_{\rm H I}$ spectrum, $\log N_{\rm H I}\la 16-$the Lyman-$\alpha$ forest regime, we consider two complementary samples that probe $z<0.5$ \citep{lehner07} and $0.5<z<2.0$ \citep{janknecht06}. We also complement the lower redshift interval with the 3C\,273 sightline analyzed by \citet{williger10}. At $z<0.5$, the data come from the high resolution STIS E140M echelle mode while at higher redshift the data come from  STIS E230M as well as VLT/UVES and Keck/HIRES data. The \HI\ column densities (and Doppler parameters) were derived by fitting the Lyman-$\alpha$ line (and higher Lyman series lines if present) thanks to the high resolution of these spectra. This method works well for systems with $\log N_{\rm H I}\la 15.5$ if several Lyman series lines are used \citep[e.g.,][]{lehner06} or with $\log N_{\rm H I}\la 14$ (depending on the $b$-value) if only the Lyman-$\alpha$ transition is used. For the  $z<0.5$ sample, several Lyman series lines were used when possible. For the higher redshift sample, \citet{janknecht06} also used different atomic transitions to constrain the Doppler parameter. We note that their sample include a few PLLSs and LLSs, but the \hi\ column densities of these systems often have errors in excess of 1 dex. We excluded those systems from our analysis. Using the maximum-likelihood method, we first fitted the two Lyman-$\alpha$ forest samples separately, finding no difference between these two redshift regimes. We therefore combined both samples and fitted them simultaneously. We find $\beta = 1.72 \pm 0.02$ and $\log C_{\rm HI } = 11.14$ in the $\log N_{\rm H I}$ interval $[13.2,14.5]$ (which is shown by the blue line in Figure~\ref{fig:fn}). Changing the upper bound by $+ 2$ dex and the lower bound by $+0.5$ dex gives consistent results (within $\la 1 \sigma$). However, changing the lower bound by $-0.1$ dex decreases $\beta$ by $-0.05$ (more than $2\sigma$), and $\beta$ drops even more if the lower bound decreases further. As indicated in Figure~\ref{fig:fn}, there is a turnover in the distribution at $\log N_{\rm H I} \simeq 13.2$, which is likely due to the incompleteness of the sample at these column densities.\footnote{\citet{janknecht06} typically found $\beta \simeq 1.60$--$1.64$, but they set a completeness for their sample at $\log N_{\rm H I} = 12.9$. Setting the lower bound to 12.9 dex, we found $\beta = 1.61$, a value very similar to theirs and substantially smaller than $\beta = 1.72 \pm 0.02$. Their completeness value was not justified, and based on our analysis a lower limit of 13.2 dex appears more appropriate. The signal-to-noise in 9 of the 11 sight lines (depending on the wavelength) indeed is not dissimilar from the lower redshift sample, where \citet{lehner07} showed that the completeness was $13.2$ dex based on an analysis of the column density distribution.} While the slope derived for Lyman-$\alpha$ forest is very similar to that predicted in recent cosmological simulations \citep{dave10}, the observations do not indicate an evolution of $\beta$ in this redshift regime, as inferred in the simulations. 

Finally, the column density distributions of the PLLSs, LLSs, and SLLSs ($19 \le \log N_{\rm H I}<20.3$) have so far remained largely uncategorized at $z\le 2.6$. For the SLLSs, $\tau_{\rm LLS}$ is far too large to estimate $N_{\rm HI}$ from the Lyman break, but in this regime, the amount of \HI\ is large enough that the Lyman-$\alpha$ transition produces damping wing from which  $N_{\rm HI}$ can be estimated. In our sample, Lyman-$\alpha$ is covered in just 7 sightlines when $\tau_{\rm LLS} >3.5$. In two of these cases, there is no detection of Lyman-$\alpha$, but the data were obtained from the low resolution FOS observations. In the other five cases, Lyman-$\alpha$ is observed, but in four of them, the equivalent width implies column densities around $10^{19}$ cm$^{-2}$ or less. As the spectral resolution of the data is low and line contamination is likely, we relied on other recent works to constrain $f(N_{\rm HI})$ in the SLLS regime. Specifically, we use the surveys of \citet{omeara07} and \citet{peroux03}, which include 16 SLLS  at $1.7<z<2.6$, overlapping the high redshift portion of our LLS sample and the Lyman-$\alpha$ forest samples.\footnote{ \citet{peroux05} subsequently produced a second survey of SLLS, but their redshift coverage mostly targeted higher redshifts with a negligible redshift path at $z<2.6$.} We estimate the total absorption path probed for the SLLS searches to be $ \Delta X \simeq 107 $ ($\Delta X \simeq 29$ for the P\'eroux et al. sample, and $\Delta X \simeq 78$ for the O'Meara et al. sample). The bins for display of the data were chosen so there are $\sim$5 systems per bin (see Figure~\ref{fig:fn}). For the PLLSs and LLSs,  we considered our  sample of QSO sightlines, where we reject sight lines having LLSs with only limits on the optical depth (and hence on $N_{\rm H I}$). The main effect of the removal of the limits is to increase slightly the normalization of the fit by $\sim0.1$ dex. This is too small a difference to have any impact on our result and should not impact the power law slope. This reduces our sample to 50 systems and a total absorption path $ \Delta X =156 $. In Figure~\ref{fig:fn} we show the adopted bins for $16.5 \le \log N_{\rm H I}\le 17.8$. The first bin corresponds to optical depths  in the intervals $[0.2,0.7]$, i.e. where our sample is incomplete; we treat this bin as a lower limit.  

We used the maximum-likelihood method to fit the data with a power law distribution in $f(N_{\rm HI})$ (Equation~\ref{e-fn2}). Our first attempt was to fit the LLS and SLLS simultaneously, but no adequate fit was found with a single slope $\beta$. We, therefore, fitted the LLS and SLLS separately. For the SLLS, we find $\beta = 0.8\,^{+0.3}_{-0.1}$ for $19.1 \le \log N_{\rm H I}\le 20.2$  (where the upper and lower bounds were allowed to vary by $\pm 0.1$ dex to estimate the errors). For LLS, we derived $\beta \approx 1.9 \pm 0.3$ for $17.2 \le \log N_{\rm H I}\le 17.7$ (as the $N_{\rm HI}$ interval spans only 0.5 dex, changing the upper and lower bounds by $\pm 0.1$ dex led to an unstable fit; we consider this result as tentative). We note that if we integrate $f(N_{\rm H I})$ in the intervals $\log N_{\rm HI}=[17.3,18.2]$ and $\log N_{\rm HI}=[18.2,20.2]$, with the respective $\beta$ functional forms (where we assume that each is correct to the point where they intersect at $\sim$18.2 dex, see Figure~\ref{fig:fn}), we find $l(X) \sim 0.5$, which is not too dissimilar from the results presented in Table~\ref{tab:stats} that gives $l(X) \simeq 0.3$, providing some independent support to our results. It is evident that more data are needed in the PLLS and the LLS/SLLS regimes to better discern the true shape of $f(N_{\rm H I})$ in these $N_{\rm H I}$ intervals. However, our analysis suggests that there must be an inflection point in $f(N_{\rm H I})$ in the LLS regime, and, likely, a second inflection point in order to connect the PLLS to the Lyman-$\alpha$ forest systems. We note that the $\beta = 1.7$ slope distribution fits well the \hi\ systems with $N_{\rm HI} \la 10^{16}$ cm$^{-2}$ (see Figure~\ref{fig:fn}), so the flattening should likely occur between $10^{16}$ and $10^{17}$ cm$^{-2}$. The $\log N_{\rm H I} = [17.7,18]$ interval will likely remain largely unconstrained owing to the difficulty in measuring  $N_{\rm H I}$ in this regime requiring either to fit the Lyman series lines \citep[e.g.,][]{lehner09} or to have very high quality S/N data to  discern the damping wings in the Lyman-$\alpha$ absorption.

In Figure~\ref{fig:fn} we also show one of the $f(N_{\rm H I})$ models in the local universe by \citet{corbelli02} (long-dashed orange curve; see their Figure~2 where we adjusted vertically their model to fit the DLA and SLLS distributions -- the model with $f(N_{\rm H})\propto N^{-3.3}_{\rm H}$ is shown). In their models, they investigated if the flattening of $f(N_{\rm HI})$ between the LLSs and DLAs  could be explained if $f(N_{\rm H})$ ($\rm H =$\,\hi\,$+$\,\hii) follows a single power law, while $f(N_{\rm H I})$ can deviate from a single power owing to the change of the ionization fraction as function of $N_{\rm H I}$. While the low $N_{\rm H I}$ systems are not well matched (in part because they attempted to fit data based on equivalent width measurements), the higher column density regimes are quite remarkably well reproduced. Other models explored the self-shielding effect on the $f(N_{\rm H I})$ of DLAs and LLSs using spherical isothermal gaseous halos \citep{murakami90,petitjean92,zheng02}, which yields a somewhat similar functional form. Hence photoionization  of a single power law population in $f(N_{\rm H I})$ could be the main cause for the complicated shape of the $f(N_{\rm H I})$  distribution.

In the higher redshift regime, \citet{petitjean93} also noted that a single $f(N_{\rm H I})$ over the entire $N_{\rm H I}$ regime was not statistically adequate, and, in particular, their data hinted as well to two flattenings in the column density distribution function, one in the PLLS regime and the other one in LLS/SLLS regime that they explained as transitions between the \HI\ systems to metal absorbers and between the neutral and ionized systems, respectively. The most recent study on  $f(N_{\rm H I})$ at $z\sim$3.7, by \citet{prochaska10} suggests an even more complicated 
$f(N_{\rm H I})$ distribution. We show in Figure~\ref{fig:fn} their $f(N_{\rm H I})$ distribution over the same range of \hi\ column density. We emphasize that while the Lyman-$\alpha$ forest (up to $\log N_{\rm H I} \la 14$), SLLS, DLA, and to a lesser extent LLS distributions are relatively well constrained, the PLLSs and \HI\ interval $14 \la \log N_{\rm H I} \la 16$ are not (see their Figure~14 for the amplitude of possible $f(N_{\rm H I})$ in each $N_{\rm H I}$ region). As already mentioned above, there appears to be no evolution in the DLA portion of $f(N_{\rm H I})$ with redshift, and a steeper slope than found by \citet{rao06} seems more appropriate for connecting the DLAs and SLLSs at low-$z$. While a similar flattening is observed in the SLLS regime,  in the low-$z$ universe  $f(N_{\rm H I})$ appears (tentatively) even flatter. A larger sample of SLLSs will be needed to confirm this as other explanations (e.g., an evolving normalization at different mean redshift or the presence of another inflection point) could account for the observed behavior. 

In  the lower $N_{\rm H I}$ regime,  $f(N_{\rm H I})$  appears to evolve from the high to  low-$z$ universe. At $\log N_{\rm H I} \le 14.5$, where  $f(N_{\rm H I})$ is well constrained at both low and high $z$, the slope becomes steeper as $z$ decreases and there is a drop in the number of systems with redshift. Without a steep decline in the UV background flux (stemming from a drop of the number of QSOs at low $z$), the number of systems would be predicted to be much lower at low $z$, suggesting that the changes in the UV background may be the dominant reason for the evolution of the Lyman-$\alpha$ forest \citep[e.g.,][]{theuns02}. Numerical simulations of a cold dark matter universe with a photoionized background dominated by the QSO light can, indeed, reproduce these properties to some extents \citep[e.g.,][]{theuns02,dave10}, but the observed evolution rate of $\beta$ is smaller than predicted. Part of the discrepancy between the models and observations could be due to the models ignoring the galactic contribution to the UV background, or more generally to an uncertainty in the strength and shape of the UV background. Large-scale galactic outflows could be thought as another uncertainty because (in the regime where $\log N_{\rm H I}> 14$) they likely increase the \hi\ absorbing cross section via deposit of cool gas in the outermost edges of galactic halos \citep{dave10}. Cosmological simulations however, suggest that galactic feedback has little impact on  $f(N_{\rm H I})$ of the Lyman-$\alpha$ forest as they only fill a small fraction of the volume, leaving the IGM filaments unscathed \citep{theuns02a}. Hence, the possible differences seen at $14.5 \la \log N_{\rm H I} \la 19$ may occur owing to the evolution of both the UV background and galactic feedback. 
Current and future efforts to provide better statistics for the PLLSs and LLSs at both low and high-$z$ should provide direct constraints on the UV background evolution and cosmological simulations.

%%%%%%%%%%%%%%%%%%%%%%%%%%%%%%%%%%%%%%%%%%%%%%%%%%%%%%%%%%

\section{LLSs and the Gaseous Halos of Galaxies}
\label{sec:gal}

At very low redshift, the  connection between LLSs, galaxies, and large-scale 
structures has been examined for a small number of individual systems discovered using \textit{HST} and the \textit{Far 
Ultraviolet Spectroscopic Explorer} (\textit{FUSE}). These studies have found LLSs associated with 
individual galaxies ($0.2 L_* \la L \la 3.4L_* $) at impact parameters $\rho\sim30-100$ kpc
 \citep{cp00,jenkins03,tripp05,cooksey08,lehner09}. Some low redshift 
LLSs are metal enriched \citep[i.e., $Z \gtrsim 0.3 Z_{\sun}$, e.g.,][]{cp00,prochaska06b,lehner09} while some are relatively metal-poor 
\citep[i.e., $Z \lesssim 0.1 Z_{\sun}$, e.g.,][]{pb99,cooksey08,zonak04}. The presence of metal-enriched material far from the 
central star forming regions of galaxies suggests some LLSs are sensitive to the nature of 
feedback in galaxies. The existence of extremely metal-poor systems suggests the gas 
probed by some LLS absorption originates outside of galaxies, perhaps tracing IGM matter falling onto a galaxy. An example of a LLS tracing very low metallicity
($Z \sim 0.02 Z_{\sun}$) gas falling onto a near solar, 0.3$L_*$ galaxy at $z \sim0.274$ will be described in \citet{ribaudo11}.
In addition to the observational evidence, numerical simulations also predict a physical association
of LLSs with the gravitational potential of galaxies. These simulations show LLSs arising from infalling streams of intergalactic gas 
as well as outflowing gas ejected from galaxies due to stellar feedback
 (Gardner et al. 2001; Dekel \& Birnboim 2006; Kohler \& Gnedin 2007; Keres et al. 2009; Kacprzak et al. 2010, Fumagalli et al. 2011, Stewart et al. 2011; 
but also see, Mo \& Miralda-Escud$\acute{e}$ 1996; Maller et al. 2003).

\begin{figure*}
 \epsscale{1.0}
 \plotone{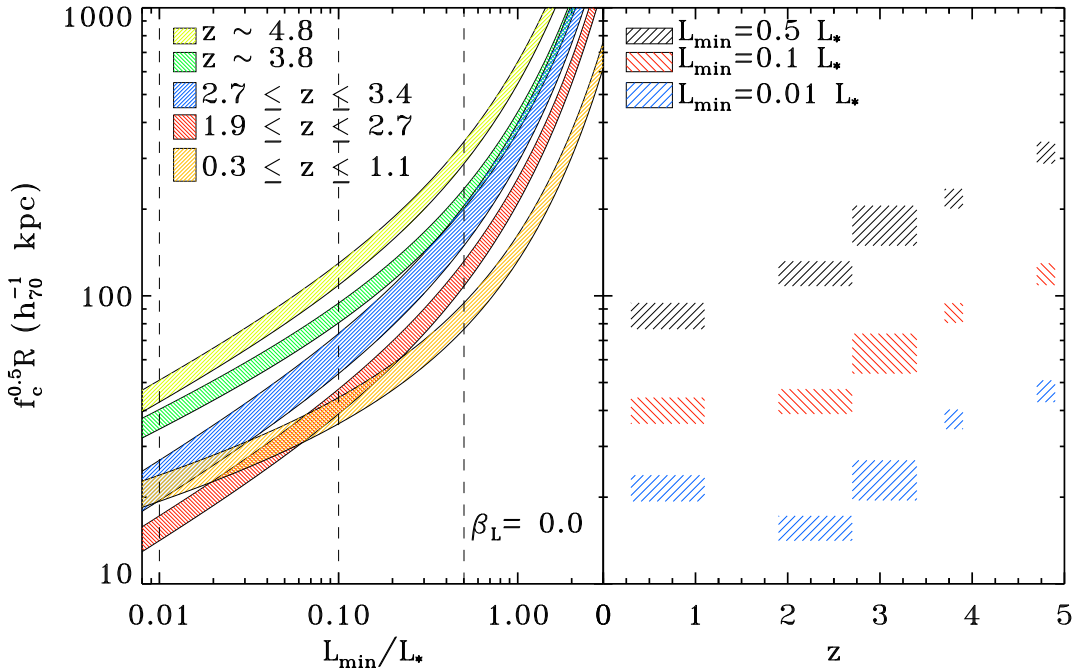}
 \caption{ The left panel shows the statistical absorption distance ($f_c^{0.5} R$)  plotted against  $L_{\rm min}/L_*$, where $L_{\rm min}$ is 
the assumed minimum luminosity for a galaxy to give rise to LLS
absorption (see Equation~\ref{eqn:Rs}). The shaded areas represent the absorption distance for specific redshift ranges, each with
different luminosity function parameters.  The orange region corresponds to the average $\alpha$ and $\Phi_*$ values determined by \citet{faber07} using
DEEP2 over the redshift range $0.3 \la z \la 1.1$. The red and blue regions correspond to the analysis of \citet{rs09} over the redshift range $1.9 \la z \la 2.7$ and
$2.7 \la z \la 3.4$, respectively. Lastly, the green and yellow regions correspond to the analysis of \citet{vanderburg10} using the CFHT Legacy Survey Deep fields at redshifts
$z \sim 3.8$ and $z \sim 4.8$. The right panel shows the evolution of the statistical absorption distance as a function of redshift. The colored shaded regions correspond to a 
minimum luminosity cut designated in the left panel with the dashed lines. The width of the regions corresponds to the redshift range probed by the 
particular survey used to calculate the galaxy luminosity function
and the height of the regions correspond to the range in $R_s$ seen as we vary the redshift over the acceptable range. Note the evolution in the physical size of absorbers 
in the right panel does not imply single galaxies are evolving on the same timescale. Rather, the right panel implies the physical size of the gaseous envelope of a mean galaxy 
at a specific redshift undergoes significant evolution over cosmic time.}
 \label{fig:galx}
 \end{figure*}

Based on these observational and theoretical studies, LLSs appear to be associated with circumgalactic environments. With this knowledge, we can calculate the characteristic 
sizes of such gaseous galactic envelopes using our survey of LLSs and knowledge of the galaxy population with which they are associated. 
We rewrite $l(X) \varpropto n_{\rm LLS}\sigma_{\rm LLS}$  as:
\begin{equation}
\label{eqn:gals}
l(X)_{\rm LLS}\varpropto n_{\rm GAL}\sigma_{\rm GAL},
\end{equation}
where $n_{\rm GAL}$ is the comoving number density of galaxies giving rise to LLS absorption 
and $\sigma_{\rm GAL}$ is the projected physical cross section of galaxies to columns $\log N_{\rm HI} \ge 17.5$ (for comparison with R2,RP10 samples).
 The comoving
number density of galaxies at a given redshift is calculated from the integration of an observationally constrained 
galaxy luminosity function. We investigate the size of absorbers assuming only galaxies with $L\ge L_{\rm min}$
give rise to LLS absorption. Thus, following  \citet{tytler87}, we rewrite Equation~\ref{eqn:gals} as:
\begin{equation}
\label{eqn:lgals}
l(X)=\frac{c}{H_0}\int_{L_{\rm min}}^{\infty} f_c \pi R^2(L) \Phi(L) dL,
\end{equation}
where $f_c \pi R^2(L)$ is the cross section for absorption, $\sigma_{\rm GAL}$, with a covering 
factor $f_c$ and  $\Phi(L) dL = \Phi_* (L/L_*)^\alpha \exp(-L/L_*) d(L/L_*)$
 is the assumed form of the galaxy luminosity function \citep{schechter76}. The comoving number density of 
galaxies that contribute to 
the LLS population is determined by our choice of $L_{\rm min}$, and we use all galaxies with $L\ge L_{\rm min}$ in this estimation of the mean $\sigma_{\rm GAL}$. 
We note that several previous treatments of the
gaseous halos around galaxies have allowed for a Holmberg-like scaling of the physical extent of the gas with 
$R(L) = R_*(L/L_*)^{\beta_{\rm L}}$, where  $R_*$ is the projected radial extent of the absorbing gas 
associated with an $L_*$ galaxy \citep{tytler87}.
Numerous galaxy-absorber studies have shown if the radial extent of galaxies is allowed to scale with luminosity, $R(L)$ serves as the
effective cutoff for observed absorption out to that projected distance
\citep[i.e.,][]{kacprzak10,chen10,kacprzak08,chen01}. However, we are considering the physical extent of absorbing gas averaged
over all galaxy types
and sizes, with our only selection criterion being $L\ge L_{\rm min}$, and over a very wide range in redshift.  
Over time, the galaxies giving rise to LLSs may be best described with
an evolving $\beta_L$, but as we are generalizing our analysis to the size of the gaseous envelope around a \textquotedblleft mean\textquotedblright galaxy, 
averaged over all morphologies, star formation properties, sizes, etc., we adobt $\beta_L=0.0$. Our results therefore describe the mean extent of circumgalactic 
gas about galaxies $L\ge L_{\rm min}$. 
  
Equation~\ref{eqn:lgals} can be solved using the incomplete $\Gamma$ function, giving the statistical absorption radius of a galaxy
\begin{equation}
\label{eqn:Rs}
R_s=f_c^{0.5}R=\left[ \frac{c \pi \Phi_*}{H_0 l(X)}\Gamma \left( 2\beta_{\rm L}+\alpha+1,\frac{L_{\rm min}}{L_*} \right) \right]^{-0.5}.
\end{equation}
The radius $R_s$ is therefore the mean radial extent of gas about an average host galaxy scaled by $f_c^{0.5}$, while $\pi R_s^2=\sigma_{\rm GAL}$ 
is the projected area of such a galaxy for which $\log N_{\rm HI} \ge 17.5$ for a given choice of $L_{\rm min}$. 
We plot $R_s$ in Figure~\ref{fig:galx} as a function of the assumed $L_{\rm min}$ in the left panel and  redshift in the right panel. 
In the left panel, the shaded regions correspond to different luminosity function parameters, which are appropriate 
for the redshift ranges given in the legend. The luminosity function parameters are 
observationally determined and restricted to the redshift range probed by each survey. 
The right panel shows the statistical absorption radius as a function of redshift for three snapshots of $L_{\rm min}$.
The width of the shaded regions is determined from the redshift range
of the survey used to calculate the luminosity function. The height of each region spans the $R_s$ value 
predicted for the range in redshift. 
 The recent study of \mgii\ absorbers and galaxies at $z < 0.5$ by \citet{chen10} found $f_c=0.70$ for 
the strongest \mgii\ absorbers out to $R_*=75$ kpc (with $\beta_L=0.35$). The introduction of a non-unity covering factor will thus increase the 
values for $R$ by $\sim$10--30\% compared with $R_s$. 

From Figure~\ref{fig:galx}, we can draw several inferences about the evolution and properties of the 
galactic environments giving rise to LLS absorption, albeit with some limitations.  
We are describing the mean extent of $\tau \geq 2$ \hi\ gas with no assumptions about which galaxies give rise to the absorption. 
The evolution in $R_s$ does not track the evolution of individual galaxies, only the mean galaxy with $L \geq  L_{\rm min}$ for each $z$. 
Any change in the physical cross section for the mean galaxy at each redshift does not imply individual galaxies are 
evolving on that timescale, as it is likely the case the galaxies giving rise to LLSs at $z\sim5$ are not the same galaxies 
giving rise to LLSs at $z\sim1$. With these limitations in mind, several inferences can be drawn from this approach.

The $L \ge L_*$ galaxies alone cannot account for the
observed population of LLSs, because $R_s$ would be inconsistent with previous galaxy-absorber observations, especially 
at $z\ge2$ where the sizes implied for LLSs would be quite large compared with 
observations \citep{steidel10}. Extending the integration of Equation~\ref{eqn:lgals} to sub-$L_*$ galaxies 
produces $R_s$ values more consistent with the impact parameters found independently by 
other studies \citep[e.g.][]{bouche07,kacprzak10,chen10}. It is not clear how small $L_{\rm min}$ 
should be before we can account for the entire population of LLSs, but Figure~\ref{fig:galx} highlights the importance and 
need for deep observations of QSO fields to confidently relate absorbers
to specific galaxies. This conclusion is not surprising as \mgii\ studies and individual LLS observations show 
sub-$L_*$ ($0.25 \la L/L_* \la 0.76$) galaxies contribute to the population of optically thick
absorbers \citep[e.g.][]{steidel10,kacprzak10,chen10,lehner09,kacprzak08}. However, our analysis suggests the less luminous galaxy population 
may be the dominant source of LLSs. A similar scenario has been suggested for \mgii\ absorbers over the 
redshifts $0.37 \le z \le 0.82$, where \citet{caler10} find evidence that at least 
$70$--$75$\% of the \mgii\ absorber host galaxies are fainter than $0.56L_*$.

Figure~\ref{fig:galx} also highlights a significant evolution in the
physical cross section of the mean absorbing galaxy as a function of
redshift. For $L_{\rm min} \sim 0.1L_*$, $R_s$ decreases by a factor
of $\sim$3 from $z\sim$5 to 2, but remains relatively constant from
$z\sim$2 to 0.3. This is remarkable as it suggests the physical cross
section of the gaseous envelopes of a mean galaxy has decreased
significantly over a very short epoch, but for the majority of cosmic
time the physical extent of gas about a mean galaxy has been fairly
constant.  This relatively constant nature of absorption cross section
at low-$z$ was also noted by \citet{nestor05}, who found evidence for
little evolution in the physical size of \mgii\ absorbers as a
function of redshift over $0.3 \le z \le 1.2$ (for $0.001 \le L_{\rm
min}/L_* \le 0.25 $).

Changes in physical cross section can be brought on by evolution in
the typical covering factor as well as typical radial extent.
However, changes in $f_c$ alone likely cannot be responsible for the
large drop in the physical cross section of the mean galaxy given the
typical values observed at low redshift.  While a change in the
typical radial extent, $R$, is a likely cause, other factors could
influence our perception of the cross section for the mean galaxy at a
given redshift.  Evolution in the power law index, $\beta_L$,
associated with changes in the relative fraction of high versus low
luminosity galaxies giving rise to LLSs could alter the mean cross
section calculated here.  For example, if at high redshifts ($z\sim5$)
the majority of LLSs arise in the circumgalactic gas of relatively
high mass, bright galaxies, but at low redshifts ($z\sim0.3-1$) the
majority of LLSs arise in the environments of low mass, relatively low
luminosity galaxies, we would expect an evolution in the mean physical
cross section of LLS absorption similar to what is shown in
Figure~\ref{fig:galx}.  An evolution in $L_{\rm min}$ with redshift
would have an affect similar to an evolving $\beta_L$.

As we alluded to above, 
there are two commonly invoked scenarios for producing circumgalactic gas at such large distances 
from the central regions of galaxies. In the first, galactic-scale outflows drive gas to large radial 
distances from the main body of a galaxy providing for \mgii\ and LLS absorption \citep{bouche06}. Evidence for this
has been presented by \citet{bouche07}, who found starburst galaxies within 50 kpc for $\sim70\%$ of a sample of strong \mgii\ absorbers. 
\citet{prochter06} have also argued the importance of outflows to \mgii\ selected systems 
based on the similarity in the evolution in the redshift incidence of strong \mgii\ absorbers and the star formation rate 
density of the Universe for $z<2$. Combined with constraints on the size of the 
galaxies giving rise to the \mgii\ absorption, this suggests such systems are produced through feedback processes in 
low mass galactic halos. In addition, other recent works have connected \mgii\ selected absorbers to galactic outflows
 at $z\sim0.7$ \citep{nestor10}, $0.5 < z < 1.4$ \citep{menard09},
and $2 \lesssim z \lesssim 3$ \citep{steidel10}. 

The second scenario assumes much of the circumgalactic material traced by LLSs 
 is intergalactic gas being accreted onto the galaxies. To maintain the low apparent ionization 
conditions of LLSs \citep{lehner09,cooksey08}, the gas should not be shock heated as it is accreted. Such low-ionization gas falls under
the phenomenon of 
cold mode accretion (CMA) \citep{keres05} predicted to be directed along the filamentary structure of
the Universe, allowing galaxies to draw gas from large distances. CMA can account for the observational properties of 
galaxies inconsistent with the traditional shock-heated accretion models, such as 
the color bimodality of galaxies and the  decline 
of the cosmic star formation rate at low redshifts \citep{keres05,db06,dekel09a,dekel09b,keres09}. 
Support for CMA has been suggested in recent studies of \mgii\ absorbers where no correlation between 
the \mgii\ 
absorption strength and galaxy color was found, indicating the origin of the absorbers is not tied to the 
star formation history of the associated galaxy \citep{chen10}. \citet{chen10} conclude \mgii\ absorbers (and LLSs as an 
extension) are a generic feature of galaxy
environments and that the gas probed by \mgii\ absorption is likely intergalactic in origin. There is more direct observational evidence to 
support this origin for some LLSs. The nearly primordial LLS detected by \citet{ribaudo11} within 40 kpc of a near solar galaxy is similar in 
ionization state and metallicity
to the low-metallicity absorbers reported in \citet{cooksey08} and \citet{zonak04}.

While outflows and infall must play an important role in the composition and maintenance of circumgalactic environments,  observations of
a few systems suggest gas ejected to large distances during galaxy
mergers and tidal interactions could also be responsible for some of the
observed LLSs \citep[e.g.,][]{jenkins03,lehner09}.  Other studies have
suggested the high velocity clouds (HVCs) seen about the Milky Way may
be analogs for the higher redshift LLSs or \mgii\ systems
\citep{charlton00,richter09,stocke10}.  In the Milky Way and the
nearby Magellanic Clouds, the HVCs probe outflows related to galactic
fountains and winds \citep{keeney06, zech08, lehner07a, lehner09b}, the
infall of low-metallicity gas \citep[e.g.,][]{wakker01, wakker08,
 thom08}, and the tidal debris stripped from the Magellanic Clouds
(and others) as they interact with each other and the Milky Way
\citep[e.g.,][]{putman03}.  Thus, these potential LLS analogs have a
wide range of origins, although many of the Milky Way HVCs tend to
reside at much smaller impact parameters than suggested for the LLSs \citep[$\rho \leq 10-20$ kpc][]{lehner10,wakker08, thom08}.  
On the
other hand tidal remnants from galactic interactions or gas outflows
from its satellites are about 50--100 kpc
from the Milky Way. These local analogs underline the complex task in
defining what kind of phenomena the LLSs trace and if one dominates
over the others.

Discriminating between these scenarios using only the correlations of
redshifts and equivalent widths of the \mgii\ lines with other
parameters has been difficult.  The availability of \hi\ column
density information for a large number of LLSs offers a path to
studying the metallicities of the LLSs/\mgii\ systems at low
redshifts.  Further studies specifically targeting the galactic
environments of LLSs, where the metallicities of the
LLSs and the galaxies can be compared, will be critical to further
characterize the nature of the absorbers and the role these systems
play in the movement of gas into and out of the halos of
galaxies.  With metallicity playing a fundamental role in 
discriminating between these two scenarios \citep[e.g.,][]{fumagalli11},
absorbers will need to be selected based on \hi\ absorption to provide
a comprehensive picture of the nature and orgin of circumgalactic gas.

%%%%%%%%%%%%%%%%%%%%%%%%%%%%%%%%%%%%%%%%%%%%%%%%%%%%%%%%%%%%%%%%%%%%%%%

\section{Summary and Concluding Remarks}
\label{sec:sum}

Using FOS and STIS {\it HST}\ archival observations, we have compiled the largest sample of QSOs to date with coverage of the Lyman limit over the redshift range  $0.24 \le z \le 2.59$. 
We have used these observations to study the population of LLSs over these redshifts. 
In considering candidates for our R1 (R2) sample, we included only the data from objects where the spectral 
quality was judged to be sufficient to reliably detect a LLS with $\tau_{\rm LLS} \ge 1$ ($\tau_{\rm LLS} \ge 2$). The sample R1 (R2) contains
229 (249) QSOs, covering a total redshift path of $\Delta z=79$ $(96)$ and a total of 61 (50) LLSs. This marks a factor of $\sim3-4$ increase in the number of LLSs and redshift path sampled
over the most up-to-date work by \citet{stengler95} and \citet{jannuzi98} in this redshift regime.  In addition to our statistical sample, we have catalogued 206 
low redshift LLSs from the FOS and STIS archives, which increases the sample of LLSs by a factor of $\sim$10 for the $ z \le 2.6$ sample. The robustness of our samples allowed 
us to examine the evolution  of LLSs over $0.24 \leq z \leq 2.59$ for the R1 and R2 samples and from $0.2\la z \la 5.0$ for the RP10 sample that combines our R2 sample with the high redshift sample of \citet{prochaska10}. Our main results are as follows:

\begin{enumerate}

\item We find the redshift density to be well fitted by the power law $l(z) \propto (1+z)^\gamma$ (Equation~\ref{eqn:nz}). We find for sample R1 (R2) $\gamma=$~\gammaone\ (\gammatwo). For the RP10 sample at $z \le 5$, $l(z)$ is well modeled by a single power law with $\gamma =$~\gammac\ (for $\tau_{\rm LLS} \ge 2$). 

\item Assuming a standard $\Lambda$CDM cosmology with our RP10 sample, we find $l(X)$, which is proportional to the product of the comoving 
number density of absorbers, $n_{\rm LLS}$, and the average physical size of an absorber, $\sigma_{\rm LLS}$, decreases by a factor 1.5 from $z\sim$5 to 3.  The evolution of $l(X)$ at $z \le 2.6$ 
has slowed considerably, decreasing by a similar factor for $z\sim$2.6 to 0.25. This indicates the environments  which give rise to LLSs experienced dramatic changes in the first $\sim$2 Gyr after $z\sim$5, then more slowly evolved over the following $\sim$8 Gyr.

\item We calculate the average proper distance, $\Delta r_{\rm LLS}$, a photon travels before encountering a $\tau_{\rm LLS} \ge 2$ LLS and compare this result with the predicted mean free path of hydrogen ionizing photons. The ratio of  $\Delta r_{\rm LLS}$ and the mean free path from $z\sim$5 to 0 suggests the $\tau_{\rm LLS} < 2$ absorption systems have become increasingly more important for absorption of Lyman continuum photons as the Universe 
has evolved.

\item We model the column density distribution function, $f(N_{\rm HI})$, for the various $N_{\rm HI}$ regimes at $z \le 2.6$ using a functional form $f(N_{\rm HI}) \propto N^{-\beta}_{\rm HI}$. We show that a single power law cannot fit the entire observed $N_{\rm HI}$ regime. Instead several slopes are needed. For the LLSs, we derive $\beta \sim$1.9. The functional form in the Lyman-$\alpha$ forest regime ($\beta = 1.72 \pm 0.02$) and
in the SLLS regime ($\beta \sim$0.8) suggests the distribution has two inflection points. For the DLA regime, $\beta \sim$1.8 seems appropriate for connecting $f(N_{\rm HI})$ between the DLAs and SLLSs. Simple models assuming a single power law in $f(N_{\rm H})$ with absorbers photoionized by the UV background reproduce the $f(N_{\rm HI})$ distribution remarkably well.

\item We observe little redshift evolution in $f(N_{\rm HI})$ for the SLLSs and DLAs from high ($z\sim$3.7) to low ($z\le 2.6$) redshifts. However, there is evidence that $f(N_{\rm HI})$ evolves from high to low redshift at $\log N_{\rm HI}\la 17.7$, which coincides with the strong evolution seen in the UV background and star-formation rates of galaxies over similar redshifts.  

\item Assuming LLSs arise in circumgalactic gas, we find the physical cross section of the mean galaxy  at each redshift to LLS absorption decreased by a factor of $\sim$9 from $z\sim$5 to 2 and subsequently stayed relatively constant.  We argue sub-$L_*$ galaxies must contribute significantly to the absorber population.

\end{enumerate}

\acknowledgments
The authors wish to thank J.X. Prochaska who kindly made their data available for comparison with this sample prior to publication. We would also like to thank the referee for 
useful and insightful comments, as well as J.X. Prochaska and J. O'Meara for
their valuable comments. Support for this research was provided by NASA through grant HST-AR-11762.01-A from the Space Telescope Science Institute, which is operated by the Association of Universities for Research in Astronomy, Incorporated, under NASA contract NAS5-26555. Further support comes from NASA grant NNX08AJ31G. This research has made use of the NASA Astrophysics Data System Abstract Service and the Centre de Donn\'ees de Strasbourg (CDS).

%%%%%%%%%%%%%%%%%%%%%%%%%%%%%%%%%%%%%%%%%%%%%%%%%%%%%%%%%%%%%%%%%%%%%%
%%
%% References:

\clearpage

%%%%%%%%%%%%%%%%%%%%%%%%%%%%%%%%%%%%%%%%%%%%%%%%%%%%%%%%%%%%%%%%%%%%%%
%%%%%  TABLES:

% %%%%%%%%%%%%%%%%%%%%%%%%%
% %%% Table 1

\begin{longtable}{llllll}
\tabletypesize{\footnotesize}
\tablewidth{0pt} 
\tablecaption{Observations from STIS archive  \label{tab:stis}}
\tablehead{ 
\colhead{Object} & \colhead{R.A. (J2000)} & \colhead{DEC. (J2000)}& \colhead{Grating} &
\colhead{Exp. (s)} & \colhead{Id} }
%
%\startdata
%
J0001+0709 & 00 01 40.580  & +07 09 54.62 & G230L & 2304 & 8569 \\
J0012--0122 & 00 12 10.880  & --01 22 07.74 & G230L & 2304 & 8569 \\
J0018--3529 & 00 18 41.430  & --35 29 04.70 & G140L & 2392 & 8287 \\
J0021+0104 & 00 21 27.886  & +01 04 20.10 & G230L & 2355 & 9382 \\
J0021+0043 & 00 21 33.264  & +00 43 00.73 & G230L & 2379 & 9382 \\
J0021--0128A & 00 21 51.800  & --01 28 34.11 & G230L & 7900 & 8126 \\
J0022--0128B & 00 22 04.890  & --01 28 48.07 & G230L & 7900 & 8126 \\
J0032--2144 & 00 32 44.690  & --21 44 22.14 & G230L & 5037 & 8225 \\
\end{longtable}

\clearpage

% %%%%%%%%%%%%%%%%%%%%%%%%%
% %%% Table 2

\begin{longtable}{lllll}
\tabletypesize{\footnotesize}
\tablewidth{0pt} 
\tablecaption{Observations from FOS archive (G160L)  \label{tab:fosl}}
\tablehead{ 
\colhead{Object} & \colhead{R.A. (J2000)} & \colhead{DEC. (J2000)}& \colhead{Exp. (s)} & \colhead{Id}}
%
%\startdata
%
\cutinhead{preCOSTAR}
J0017+8135 & 00 17 08.563  & +81 35 08.91 & 600 & 1027 \\
J0017+8135 & 00 17 08.563  & +81 35 08.91 & 25 & 1027 \\
J0017+8135 & 00 17 08.563  & +81 35 08.91 & 25 & 1027 \\
J0027+2241 & 00 27 15.400  & +22 41 58.50 & 1331 & 2424 \\
J0047+0319 & 00 47 05.900  & +03 19 54.90 & 530 & 2424 \\
J0057--2643 & 00 57 58.012  & --26 43 14.14 & 1200 & 3199 \\
J0117--0841 & 01 17 23.338  & --08 41 32.38 & 1127 & 3268 \\
J0117--0841 & 01 17 23.338  & --08 41 32.38 & 1143 & 4856 \\
\end{longtable}

\clearpage

%%%%%%%%%%%%%%%%%%%%%%%%%
%%% Table 3
\renewcommand{\thefootnote}{\alph{footnote}}
\begin{longtable}{lccccccc}
\tabcolsep=4pt
\tablewidth{0pt}
\tablecaption{List of objects used in sample R1/R2. \label{tab:Sample}}
\tabletypesize{\footnotesize}
\tablehead{
\colhead{} & 
\colhead{} &
\colhead{} &
\multicolumn{2}{c}{R1\tablenotemark{b}}&
\colhead{} &
\multicolumn{2}{c}{R2\tablenotemark{c}} \\
\cline{4-5} \cline{7-8} 
\colhead{Object\tablenotemark{a}}&
\colhead{$z_{\rm em}$}&
\colhead{}&
\colhead{$z_{\rm min}$} & 
\colhead{$z_{\rm max}$} &
\colhead{} & 
\colhead{$z_{\rm min}$} & 
\colhead{$z_{\rm max}$}
}
%
%\startdata
%
J0001+0709  & 3.234 & & 2.226 & 2.481 & & 2.214 & 2.481 \\
B0002+0507  & 1.900 & & 1.470 & 1.871 & & 1.470 & 1.871 \\
B0002--4214  & 2.760 & & 2.301 & 2.594 & & 2.301 & 2.594 \\
B0003+1553  & 0.450 & & 0.330 & 0.435 & & 0.289 & 0.435 \\
J0012--0122  & 1.998 & & 1.387 & 1.968 & & 1.387 & 1.968 \\
J0021+0043  & 1.243 & & 0.943 & 1.221 & & 0.943 & 1.221 \\
J0021+0104  & 1.829 & & 1.575 & 1.801 & & 1.575 & 1.801 \\
J0021--0128A  & 1.588 & & 1.403 & 1.562 & & 1.243 & 1.562 \\
\end{longtable}

\clearpage

%%%%%%%%%%%%%%%%%%%%%%%%%
%%% Table 4
\begin{longtable}{ lcccccc}
\tabcolsep=4pt
\tablewidth{0pt}
\tablecaption{List of LLSs. \label{tab:LLS}}
\tabletypesize{\footnotesize}
\tablehead{
\colhead{Object\tablenotemark{a}} & \colhead{$z_{\rm em}$} &
\colhead{$z_{\rm LLS}$\tablenotemark{b}} & \colhead{$\tau_{\rm LLS}$} & \colhead{$\log N_{\rm H I}$} & \colhead{Sample\tablenotemark{c}} & \colhead{Instrument\tablenotemark{d}}}
%
%\startdata
%
B0002--4214  & 2.760 & 2.301 & $> 4.06$ & $>17.81$ &  R1,R2  &  FOS-H  \\
J0012--0122  & 1.998 & 1.727 & $ 1.70\pm 0.08$ & $17.44_{- 0.02}^{+ 0.01}$ &  R1  &  STIS  \\
J0012--0122  & 1.998 & 1.387 & $> 2.33$ & $>17.57$ &  MgII  &  STIS  \\
J0021+0043  & 1.243 & 0.943 & $> 2.97$ & $>17.67$ &  MgII  &  STIS  \\
J0021+0104  & 1.829 & 1.576 & $> 2.47$ & $>17.59$ &  MgII  &  STIS  \\
J0021--0128A  & 1.588 & 1.499 & $ 0.96\pm 0.07$ & $17.19_{- 0.04}^{+ 0.03}$ &  PLLS  &  STIS  \\
J0021--0128A  & 1.588 & 1.409 & $ 1.16\pm 0.13$ & $17.27_{- 0.06}^{+ 0.04}$ &  R1  &  STIS  \\
J0021--0128A  & 1.588 & 1.243 & $> 2.01$ & $>17.50$ &  R2  &  STIS  \\
\end{longtable}

\clearpage

%%%%%%%%%%%%%%%%%%%%%%%%%%%%%%%%%%%%%%%%%%%%%%%%%%%%%%%%%%%%%%%%%%%%%%
%%
%%  Table 5
\begin{deluxetable}{ cccccccc}
\tablewidth{0pt}
\tablecaption{Redshift Density of LLSs for R1 and R2. \label{tab:stats}}
\tabletypesize{\footnotesize}
\tablehead{
\colhead{z} & \colhead{$\Delta X$} &
\colhead{$\Delta z$} & \colhead{$N$} & \colhead{$\langle z_{\rm LLS} \rangle$} & \colhead{$l(z)$} & \colhead{$l(X)$} & \colhead{$\Delta r_{\rm LLS}$ (h$_{70}^{-1}$ Mpc)}}
\startdata
\cutinhead{$\tau_{\rm LLS} \geq 1$}

$[0.255,2.594]$& 195.73& 79.23&      61& 1.434& $ 0.77\pm 0.11$& $ 0.31\pm 0.04$ & ......\\
$[0.255,1.060]$& 59.13& 29.54&      15& 0.785& $ 0.51\pm 0.14$ & $ 0.25\pm 0.07$ &         3047\\
$[1.060,1.423]$& 48.54& 19.56&      16& 1.194& $ 0.82\pm 0.22$ & $ 0.33\pm 0.09$ &         1213\\
$[1.423,1.824]$& 45.63& 16.47&      15& 1.613& $ 0.91\pm 0.25$ & $ 0.33\pm 0.09$ &          731\\
$[1.824,2.594]$& 42.44& 13.66&      15& 2.142& $ 1.10\pm 0.32$ & $ 0.35\pm 0.10$ &          392\\
\cutinhead{$\tau_{\rm LLS} \geq 2$}
$[0.242,2.594]$& 234.70& 96.14&      50& 1.515& $ 0.52\pm 0.08$& $ 0.21\pm 0.03$ & ......\\
$[0.242,1.078]$& 80.38& 40.03&      15& 0.895& $ 0.37\pm 0.10$& $ 0.19\pm 0.05$&         3645\\
$[1.078,1.544]$& 73.50& 29.03&      15& 1.157& $ 0.52\pm 0.14$& $ 0.20\pm 0.05$&         1995\\
$[1.544,1.947]$& 44.28& 15.52&      10& 1.751& $ 0.64\pm 0.21$& $ 0.23\pm 0.08$&          917\\
$[1.947,2.594]$& 36.54& 11.56&      10& 2.257& $ 0.86\pm 0.30$& $ 0.27\pm 0.09$&          457
\enddata
\end{deluxetable}

%%%%%%%%%%%%%%%%%%%%%%%%%%%%%%%%%%%%%%%%%%%%%%%%%%%%%%%%%%%%%%%%%%%%%%
%%
%%  Table 6
\begin{deluxetable}{ lcccc}
\tablewidth{0pt}
\tablecaption{Parameters of $l(z) = l_0 (1+z)^\gamma$. \label{tab:comp_stats}}
\tabletypesize{\footnotesize}
\tablehead{
\colhead{Paper} & \colhead{$\tau_{\rm LLS} \geq$} & \colhead{$z$}& \colhead{$l_0$\tablenotemark{a}} &
\colhead{$\gamma$} }

\startdata

Sargent et al. 1989 & 1 & $[0.6,3.6]$& $0.76$ &$0.68 \pm 0.54$   \\
Lanzetta 1991 & 1 & $[0.35,2.5]$ & 1.2 & $0.3 \pm 0.9$  \\
Storrie-Lombardi et al. 1994 & 1 & $[0.4,4.7]$ &  0.27 & $1.55 \pm 0.45$ \\
Stengler-Larrea et al. 1995 & 1 & $[0.3,4.2]$& 0.25 & $1.5 \pm 0.39$ \\
This Paper & 1 & $[0.25,2.59]$& 0.28\tablenotemark{a} &\gammaone\\
Prochaska et al. 2010 & 2 & $[3.5,4.4]$& 0.0006\tablenotemark{a} &$5.2 \pm 1.5$ \\
This Paper & 2 & $[0.24,2.59]$& 0.17\tablenotemark{a} &\gammatwo\\
This Paper& 2 & $[0.24,4.9]$& 0.30\tablenotemark{a} &\gammac
\enddata

\tablenotetext{a}{$l_0 =l_* (1+z_*)^{- \gamma}$}%, where $z_*=<z_{LLS}>$.}
\end{deluxetable}

%%%%%%%%%%%%%%%%%%%%%%%%%%%%%%%%%%%%%%%%%%%%%%%%%%%%%%%%%%%%%%%%%%%%%%
%%
%%  Table 7
\begin{deluxetable}{ cccccccc}
\tablewidth{0pt}
\tablecaption{Redshift Density  of LLSs for RP10. \label{tab:stats2}}
\tabletypesize{\footnotesize}
\tablehead{
\colhead{z} & \colhead{$\Delta X$} &
\colhead{$\Delta z$} & \colhead{$N$} & \colhead{$\langle z_{\rm LLS} \rangle$} & \colhead{$l(z)$} & \colhead{$l(X)$} & \colhead{$\Delta r_{\rm LLS}$ (h$_{70}^{-1}$ Mpc)}}
\startdata
$[0.242,1.078]$ & 80.38 & 40.03 &      15 & 0.895 & $ 0.37\pm 0.10$ &  $ 0.19\pm 0.05$ &        3645 \\
$[1.078,1.544]$ & 73.50 & 29.03 &      15 & 1.157 & $ 0.52\pm 0.14$ &  $ 0.20\pm 0.05$ &        1995 \\
$[1.544,1.947]$ & 44.28 & 15.52 &      10 & 1.751 & $ 0.64\pm 0.21$ &  $ 0.23\pm 0.08$ &         917 \\
$[1.947,2.594]$ & 36.54 & 11.56 &      10 & 2.257 & $ 0.86\pm 0.30$ &  $ 0.27\pm 0.09$ &         457 \\
$[3.500,3.663]$ &119.91 & 31.07 &      56 & 3.593 & $ 1.80\pm 0.26$ &  $ 0.47\pm 0.07$ &          94 \\
$[3.663,3.925]$ &109.00 & 27.62 &      55 & 3.799 & $ 1.99\pm 0.29$ &  $ 0.50\pm 0.07$ &          77 \\
$[3.925,4.907]$ & 71.98 & 17.50 &      45 & 4.125 & $ 2.57\pm 0.44$ &  $ 0.63\pm 0.11$ &          50 
\enddata
\end{deluxetable}

\end{document}